\documentclass[11pt,a4paper,iop]{emulateapj}

\usepackage{graphicx}
\usepackage{ctable}
\usepackage{multirow}
\usepackage[rightcaption]{sidecap}
\usepackage{amsmath}
\newcommand\tstrut{\rule{0pt}{2.4ex}}

\shorttitle{Exploring the diversity of groups at 0.1$<$z$<$0.8}
\shortauthors{Connelly, et al.}

\begin{document}

%% LaTeX will automatically break titles if they run longer than one line. However, you may use \\ to force a line break if you desire.
\title{Exploring the diversity of groups at 0.1$<$z$<$0.8 with X-ray and optically selected samples}

%% Use \author, \affil, and the \and command to format author and affiliation information.
%% Note that \email has replaced the old \authoremail commandfrom AASTeX v4.0. You can use \email to mark an email address
%% anywhere in the paper, not just in the front matter. As in the title, use \\ to force line breaks.
\author{J.L. Connelly}
\affil{Max-Planck-Institut f\"{u}r extraterrestrische Physik, Giessenbachstrasse 85748 Garching Germany}
%\myemail
\author{David J. Wilman\altaffilmark{1}, Alexis Finoguenov\altaffilmark{1,2}, Annie Hou\altaffilmark{3}, John S. Mulchaey\altaffilmark{4}, Sean L. McGee\altaffilmark{5,6}, Michael L. Balogh\altaffilmark{5}, Laura C. Parker\altaffilmark{3}, Roberto Saglia\altaffilmark{1}, Robert D.E. Henderson\altaffilmark{3,5}, Richard G. Bower\altaffilmark{6}}

%% Notice that each of these authors has alternate affiliations, which are identified by the \altaffilmark after each name.  Specify alternate
%% affiliation information with \altaffiltext, with one command per each affiliation.
\altaffiltext{1}{Max Planck Institute for extraterrestrial Physics, PO Box 1312, Giessenbachstr., 85741 Garching, Germany}
\altaffiltext{2}{CSST, University of Maryland, Baltimore County, 1000 Hilltop Circle, Baltimore, MD 21250, USA}
\altaffiltext{3}{Department of Physics \& Astronomy, McMaster University, Hamilton ON L8S4M1, Canada}
\altaffiltext{4}{Observatories of the Carnegie Institution, 813 Santa Barbara Street, Pasadena, California, USA}
\altaffiltext{5}{Department of Physics and Astronomy, University of Waterloo, Waterloo, Ontario, N2L 3G1, Canada}
\altaffiltext{6}{Department of Physics, University of Durham, Durham, UK, DH1 3LE}

%% Mark off your abstract in the ``abstract'' environment. In the manuscript style, abstract will output a Received/Accepted line after the
%% title and affiliation information. No date will appear since the author does not have this information. The dates will be filled in by the

\begin{abstract}
We present the global group properties of two samples of galaxy groups containing 39 high quality X-ray selected systems and 38 optically (spectroscopically) selected systems in coincident spatial regions at  0.12$<$z$<$0.79.  The total mass range of the combined sample is $\sim 10^{12} -  5 \times 10^{14}$ M$_{\odot}$.  Only nine optical systems are associable with X-ray systems.  We discuss the confusion inherent in the matching of both galaxies to extended X-ray emission and of X-ray emission to already identified optical systems.  Extensive spectroscopy has been obtained and the resultant redshift catalog and group membership are provided here.  X-ray, dynamical, and total stellar masses of the groups are also derived and presented.  We explore the effects of utilizing different centers and applying three different kinds of radial cut to our systems: a constant cut of 1\,Mpc and two r$_{\textrm{\tiny200}}$ cuts, one based on the velocity dispersion of the system and the other on the X-ray emission.  We find that an X-ray based r$_{\textrm{\tiny200}}$ results in less scatter in scaling relations and less dynamical complexity as evidenced by results of the Anderson-Darling and Dressler-Schectman tests, indicating that this radius tends to isolate the virialized part of the system.  The constant and velocity dispersion based cuts can overestimate membership and can work to inflate velocity dispersion and dynamical and stellar mass.   We find L$_{X}$-$\sigma$ and M$_{stellar}$-L$_{X}$ scaling relations for X-ray and optically selected systems are not dissimilar. The mean fraction of mass found in stars, excluding intra-cluster light, for our systems is $\sim$0.014 with a logarithmic standard deviation of 0.398 dex.  We also define and investigate a sample of groups which are X-ray underluminous given the total group stellar mass.  For these systems the fraction of stellar mass contributed by the most massive galaxy is typically lower than that found for the total population of groups implying that there may be less IGM contributed from the most massive member in these systems.  80\% of 15 underluminous groups have less than 40\% of their stellar mass in the most massive galaxy which happens in less than 1\% of cases with samples matched in stellar mass, taken from the combined group catalog.
\end{abstract}

%% Keywords should appear after the \end{abstract} command. The uncommented example has been keyed in ApJ style. See the instructions to authors
%% for the journal to which you are submitting your paper to determine what keyword punctuation is appropriate.
%% see http://authors.iop.org/atom/help.nsf/0/48B2E680B880A050802574B400529D64?OpenDocument&journalid=ApJ#_Toc1


%% In the first two sections, notice the use of the natbib \citep and \citet commands to identify citations.  The citations are
%% tied to the reference list via symbolic KEYs. The KEY corresponds to the KEY in the \bibitem in the reference list below. We have
%% chosen the first three characters of the first author's name plus the last two numeral of the year of publication as our KEY for

%SECTION:  INTRODUCTION
\section{Introduction}

The majority of galaxies in the Universe lie in galaxy groups \citep{Eke2004a}.  Over cosmic time, groups grow hierarchically by accreting individual galaxies and smaller groups from their surrounding filamentary structure; thus, they are evolving environments.  Even within limited redshift regimes, groups are observed to have diverse properties. Local studies \citep[e.g.][]{Zabludoff2000} reveal that their galaxy populations vary from being dominated by early (as in typical clusters) to late-type (as in the field population) galaxies.  They range from ``poor" groups containing a relatively small number of galaxies (commonly identified via optical selection methods) to massive systems (commonly identified via X-ray emission and weak lensing).  The typical velocity dispersion within galaxy groups is comparable to the internal velocities of the galaxies they contain, making them ideal for galaxy-galaxy mergers and interactions. Therefore, groups are both important in their own right and as the predominant environment of galaxies.

Galaxy groups are not trivial to identify.  At higher redshifts, they are most easily found via the X-ray emission of their Intra-Group Medium (IGM).  X-ray surveys are biased towards selecting groups with rich IGM, and may not be typical of the dominant group population which shapes most galaxies in the Universe.  Samples selected optically may be dominated by overdensities of galaxies not yet fully virialized.  Different physical processes are likely to be active in these two regimes and thus a comparison of groups selected via these two disparate methods can illuminate these physical phenomena.

To fully understand groups as the environment in which the majority of galaxies reside and evolve requires both a significant number of groups and significant information on the galaxy group members themselves.  Wide-field surveys such as zCOSMOS and DEEP2 have identified many galaxy groups up to redshift $\sim$1 and $\sim$1.3 respectively \citep{Lilly2009,Gerke2007}.  The large sample sizes which these types of surveys yield allow for the rigorous determination of global trends in group properties.  Evolution of low-mass galaxies appears to be accelerated in groups \citep{Iovino2010} and transformation rates such as those from late to early type galaxy morphologies and from active to passive star formation activity are more than twice that in the field \citep{Kovak2010}.  The build-up of stellar mass on the red sequence since z$\sim$1 involves L* galaxies moving to the red sequence preferentially in groups \citep{Cooper2007}.  The low sampling rate and bright magnitude limits of these surveys mean, however, that the majority of groups have only a few confirmed members and thus that individual systems can be difficult to examine in detail.  

A complementary approach to these large volume surveys involves studying a smaller but well defined and well sampled selection of groups.  The Group Environment Evolution Collaboration (GEEC) has taken this approach, defining samples at z$\sim$0.5 and recently extending studies up to a redshift of 1.  Intermediate redshift work has focused on optically selected groups and examined stellar masses, colors, morphologies, and star formation histories in these systems comparing to trends observed in the field \citep{Wilman2005a,Wilman2005b, Balogh2007,Wilman2008,McGee2008,Wilman2009,Balogh2009,McGee2011}.  Our higher redshift study involves X-ray selected systems and first results show a prominent transient population, migrating from the blue cloud to the red sequence, in these groups \citep{Balogh2011a}.

Comparing properties such as mass, X-ray luminosity and temperature, velocity dispersion, and richness via scaling relations allows us to explore the integrated properties of groups and clusters and how they relate to one another. In clusters, minimizing the scatter in these relations is a necessity in order to obtain accurate constraints on cosmological parameters.  Through large, uniform samples, these relations are now relatively well constrained  and seem to be very tight, even up to relatively high redshifts.   Although group samples of similar size and quality are only recently available, group scaling relations exhibit a much greater scatter due to both larger measurement errors and greater intrinsic scatter in group properties \citep[e.g.][]{OsmondPonman2004,Rykoff2008,Giodini2009,Balogh2011b}.   Understanding the scatter in the relations in the group regime is a key part of illuminating the physical processes at play.

In order to study groups spanning a significant mass and evolutionary range and to compare the results obtained from two of the most common group identification methods, we have defined two different samples within the same physical area, one via optical spectroscopy and the other via X-ray emission.  In \citealt{Finoguenov2009} (hereafter Paper I), we presented the X-ray observations of our fields and preliminary results for our sample of X-ray selected groups.  We have since finished an extensive spectroscopic campaign, significantly improving the sampling rate and depth of galaxies in our fields, and present here our full sample of X-ray and optically selected systems.  In addition to X-ray derived luminosities and masses, well constrained membership now allows us to measure velocity dispersions and dynamical masses, stellar masses, and to search for dynamical complexity in our groups.  In this paper we present our catalog of groups and explore these global group properties.  Future work will examine the galaxy populations of these groups and search for correlations with global properties.

In \S \ref{sec:sampledef} we describe our samples. \S \ref{sec:xmeas} describes the X-ray measurements of both optical and X-ray selected groups and \S \ref{sec:spec} details the follow-up spectroscopy of the X-ray selected systems.  NIR measurements and galaxy stellar masses are described in \S \ref{sec:nirstellmass}.  Global group properties including radial cuts, membership, and velocity dispersions are detailed in \S \ref{sec:gzmem}. X-ray and dynamical estimates of total group mass, and the total mass in stars are presented in \S \ref{sec:masses}.  Dynamical complexity is explored in our systems in \S \ref{sec:substructure} via the Dressler-Schectman (DS) and Anderson-Darling (AD) Tests. \S \ref{sec:lxsig} presents the L$_{X}$-$\sigma$ relations for our samples. The `total' group masses are compared in \S \ref{sec:totmass}.  We discuss the stellar  and baryon content of our systems in \S \ref{sec:minstars} and X-ray underluminous systems in \S \ref{sec:underlum}.  Throughout this paper we assume a cosmology H$_{0}=75$ km s$^{-1}$ Mpc$^{-1}$, $\Omega_{M}=0.3$, and $\Omega_{\Lambda}=0.7$ unless mentioned otherwise.

%SECTION:  GROUP SAMPLE DEFINITION
\section{Group Sample Definition}
\label{sec:sampledef}

\subsection{Optically (Spectroscopically) Selected Groups}

Our optical sample is selected from the Canadian Network for Observational Cosmology Field Galaxy Redshift Survey 2 (CNOC2), a survey consisting of four sky patches, roughly equally spaced in RA, with a total area of about 1.5 square degrees \citep{Carlberg1999}.  UBVR$_{\textrm{c}}$I$_{\textrm{c}}$ photometry of the patches yielded $\sim$40,000 galaxies above the survey's $R_{\textrm{c}}\simeq23.0$ limit.  Follow-up spectroscopy of these fields resulted in redshifts for over 6,000 galaxies with a completeness of 48\% down to $R_{\textrm{c}}=21.5$ \citep{Yee2000}.  Groups present in the survey were then detected as pure galaxy overdensities in redshift space.  In total, over 200 groups ranging in redshift 0.12$<$z$<$0.55 were detected \citep{Carlberg2001}.  Given the optical spectroscopic wavelength range for CNOC2 spectroscopy, the effective redshift range for the full sample corresponds to the available wavelength range of the Ca II H and K spectral features.

Complementary to the existing spectroscopy, the GEEC has built up a multiwavelength dataset, including HST-ACS, infrared, and UV imaging and X-ray data (described and utilized in the present analysis), in order to study galaxy groups in the CNOC2 fields in detail.  26 of the CNOC2 groups at 0.3$<$z$<$0.55 have been actively targeted with Magellan-LDSS2 to improve the spectroscopic completeness and depth of the sample.   392 unique LDSS2 redshifts were obtained in three of the four CNOC2 patches elevating the average completeness at the co-ordinates of the targeted groups to 74\% above a limiting magnitude of $R_{\textrm{c}}=22$ \citep{Wilman2005a}.  10 groups (six in the RA14h field, and four in the RA21h) were observed with VLT-FORS2 in June and July of 2005.  These data have recently been reduced and yielded 233 previously unknown redshifts and a magnitude limit of $R=23.2$ \citep{Henderson2010}. Throughout this paper we consider only those 38 optical groups within the regions observed by the XMM-Newton + Chandra described in the next section, ensuring that the most direct comparisons between these differently identified systems are possible.

\subsection{X-ray Selected Groups}

X-ray observations of two of the four CNOC2 fields were obtained and used to identify a comparison sample of X-ray selected galaxy groups.  These groups are selected from deep XMM-Newton + Chandra data in the RA14h and RA21h CNOC2 patches using a wavelet algorithm.   A detailed discussion of their definition can be found in \cite{Finoguenov2009} but a brief overview follows.   Note that since the publication of that paper, additional XMM-Newton data (OBSIDS 0603590101 and 060359020) of the RA21h field were acquired and are included here. The total area covered by the X-ray observations was 0.2 and 0.3 square degrees for the RA14h and RA21h fields, respectively.  The total XMM exposure time for the RA21h field was  271.46 ksec and the Chandra exposure time in this field 101.88 ksec.  In the RA14h field, a total exposure time of 210.40 ksec with XMM and 89.02 ksec with Chandra were obtained.

For data from both instruments and patches, careful background and point source removal were performed and images co-added by normalizing each to account for the different sensitivity of the instruments to produce a joint exposure map.  The combined maximum effective exposure times in units of the equivalent Chandra exposure at the center of the field are 691 ksec for the RA21h patch and 469 ksec for the RA14h patch.  Wavelet reconstructions of the signal-to-noise images for the co-added image and separate XMM and Chandra images were then produced and extended source detection carried out at 32$\arcsec$ and 64$\arcsec$ spatial scales.  The positional uncertainty for the X-ray centers is of order of 10$\arcsec$ but can reach 30$\arcsec$ for low significance sources. The total number of secure detections in the RA14h and RA21h patches is 31 and 33, respectively.  An additional five sources with low significance (significance $<2$) are detected in each field.  

%SECTION:  X-RAY MEASUREMENTS
\section{X-ray Measurements}
\label{sec:xmeas}

\subsection{Fluxes and Luminosities}

For X-ray selected systems, the X-ray flux is measured within an area defined via a wavelet reconstruction of the X-ray images to optimize the S/N for the source.  Using the central positions and extents of the detections from the wavelet reconstructions, flux measurement was performed on the background and point source subtracted images.  All X-ray sources have a wavelet detection $\geq$4$\sigma$, corresponding to a certain detection in flux, but in cases where the aperture has been reduced in order to prevent merging of adjacent sources, the final measurement of significance may fall between 1 and 2.  To ensure that we only include robust X-ray measurements, we choose to include X-ray derived properties only for X-ray systems having a significance $\geq$ 1 in our analysis.   Note that negative values of significance reflect that the measured flux is lower than the average background level.  Statistical background removal sets the mean of the background to zero, while the statistical spread of actually observed counts results in a distribution around zero.

\begin{figure}[!h]
       \includegraphics[scale=1,trim=17 5 0 20,clip]{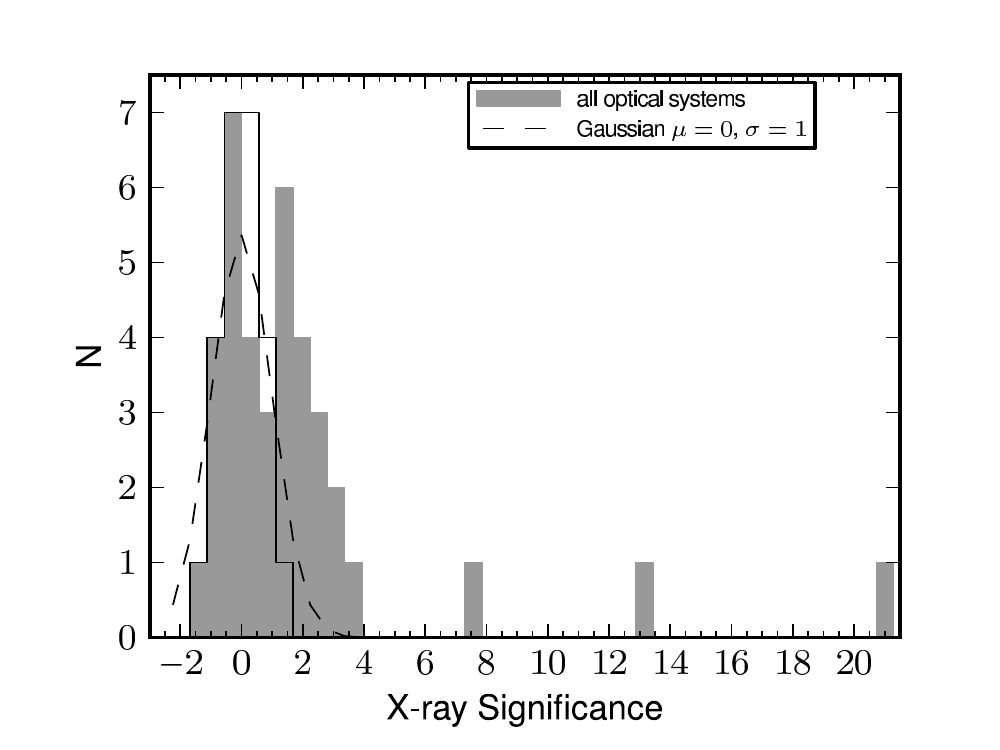}
       \caption{X-ray significance of optically selected systems. Black solid line indicates the negative portion of the histogram and its reflection above zero while the black dashed line shows a Gaussian approximation having a mean, $\mu$, of 0 and variance, $\sigma$, of 1.}
      \label{fig:fig1_Nsig}
\end{figure}

In order to measure the X-ray flux of our optical systems, we define a constant circular aperture, with a radius of 0.5$\arcmin$, surrounding the R-band luminosity weighted center of the group members.  This aperture is derived from the confusion limit in the X-ray imaging.  The distribution of the S/N of the X-ray flux estimate (hereafter X-ray significance) estimated using the residual, background and point-source removed, image of our optical systems can be seen in Fig.\,\ref{fig:fig1_Nsig}.  The histogram is double-peaked, with the first peak resulting from the noise in our flux measurements and the second the `real' peak of the X-ray significant systems. The noise peak can be approximated by a Gaussian having a mean of 0 and variance of 1 (shown as a dashed line).  The solid line indicates the negative portion of the histogram and its reflection above zero.  Comparison of this to the dashed, Gaussian approximation, line shows that the noise is slightly over-estimated, and thus the significance underestimated.  To help ensure the X-ray emission is real, we choose to include X-ray derived properties only for optical systems having a significance $\geq2$ in our analysis.  

%PLOT - X SIGNIF OF OPT GROUPS

The flux measurement is performed on background and point source subtracted XMM images only.  The total flux for each group in the 0.5$-$2 keV band is computed as in Eq.\,1 of \cite{Finoguenov2007}, extrapolating the surface brightness to r$_{\textrm{\tiny500}}$ ($\sim$ 0.6r$_{\textrm{\tiny200}}$).  The total flux and corresponding r$_{\textrm{\tiny200,X}}$ are derived iteratively from the corrected observed flux using the redshift of the group (see \S \ref{sec:xz}) and appropriate scaling relations.  \S 5.1 of \cite{Finoguenov2007} details these scaling relations.  This includes extrapolation of the measured flux, assuming the surface brightness profile measured for local groups. For X-ray and optical systems with low X-ray significance, we calculate upper limits for the X-ray flux, radius, luminosity and mass and these measures are demarcated by grey points in the appropriate relations.  Note that we are unable to robustly measure X-ray temperatures for our systems given the depth of our data.

The rest-frame X-ray luminosity in the 0.1$-$2.4 keV band is calculated by k-correcting the flux measurement within r$_{\textrm{\tiny200,X}}$ as described in \S 5.1 of \citet{Finoguenov2007} and is shown for our systems as a function of redshift in Fig.\,\ref{fig:fig2_lxz}.  We can see from this figure and Fig.\,\ref{fig:fig3_sigz} that, with increasing redshift, we are biased towards systems with higher X-ray luminosities and velocity dispersions.

\subsection{Apertures}

To ensure that the assignment of a \textit{constant} aperture centered on the luminosity weighted group center for optical groups and a \textit{variable} aperture centered on the X-ray emission peak for  X-ray systems does not significantly bias our comparison, we first compare the flux measurements for groups which are independently detected both as optical and X-ray systems.  There are only four such, relatively bright, systems having X-ray significance greater than our cut off and which are not located near the X-ray and/or optical survey edges where luminosity weighted centers can be particularly inaccurate and X-ray apertures incomplete.  All four have higher X-ray fluxes for the X-ray selected system than for the optical counterpart.  On average, the difference is a factor of two, indicating that our choice of the 0.5$\arcmin$ fixed aperture may lead to an underestimated flux despite the correction.  In all these cases, the X-ray defined aperture is larger (up to a factor of two) than the 0.5$\arcmin$.  As the larger aperture is resulting in a larger flux this indicates that our systems likely have a flatter surface brightness profile than the applied assumed relation derived from local groups.

We then test the effect of using a constant aperture for all X-ray selected systems, measuring the flux using the fixed 0.5$\arcmin$ aperture but centered on the X-ray emission peak and calculating the difference between this measurement and the flux measured using the variable aperture.  The average resultant percentage change in the flux measures is less than 4\% with measures scattered within a factor of two in both directions.   We further find that  X-ray flux measures can be greatly affected by the emission of neighboring groups.  Since this `confusion' may bias measurement of flux, we flag systems which lie in crowded X-ray regions (see \S \ref{sec:gzmem}).
  
We conclude that for individual X-ray systems, the use of the 0.5$\arcmin$ aperture may provide results more comparable to the optical systems but that, on average, the use of the X-ray aperture produces measures which are comparable to those resulting from a constant aperture but with higher S/N and less contamination.  We thus choose to use this variable aperture for our X-ray systems.  Fig.\,\ref{fig:fig2_lxz} shows the X-ray luminosity (see \S \ref{sec:xmeas}) as a function of redshift for all systems and includes for the X-ray systems luminosities measured with both choices of aperture.  The overall relation is very similar regardless of the choice of aperture.  For the brightest X-ray systems, the use of the fixed aperture usually results in a noticeably lower L$_{X}$.  This again indicates the applied local relation has a steeper surface brightness profile than our systems and implies that feedback may be more important in groups at higher redshift. This echoes the conclusion reached using the four matched systems; that  smaller X-ray apertures can lead to an under-estimation of flux for bright systems and produces greater uncertainty in X-ray derived properties.

%SECTION:  FOLLOW-UP SPECTROSCOPY
\section{Follow-up Spectroscopy}
\label{sec:spec}

Although spectroscopic completeness is relatively high in areas containing most of our optically selected groups, the extended X-ray sources (our X-ray selected systems) are often located in regions with very few previously determined redshifts.  In many cases, a system redshift was impossible to determine from the available spectroscopy.  A program of targeted follow-up spectroscopy for the X-ray detected systems was executed primarily using the VLT-FORS2 and Magellan-IMACS spectrographs.  Objects brighter than R$\approx$22 and those close to the center of the X-ray contours were preferentially targetted.  Some additional Gemini-GMOS spectroscopy has also been acquired as part of a program to extend this type of group study to higher redshift \citep{Balogh2011a}.  A summary of the follow-up spectroscopy can be found in Tab.\,\ref{tab:spec} and includes for each instrument the wavelength range and radius of the FOV and the total number of masks, spectra, and redshifts for each field.  In total 1,946 previously unknown, secure redshifts have been measured, yielding a full sample of nearly 5,000 redshifts in the RA14h and RA21h CNOC2 fields.  We provide a sample of redshifts used in this analysis in Tab.\,\ref{Schectman} while the full catalog is available in the electronic version of this article.

\subsection{FORS2 Observations}

FORS2 observations were conducted over the course of three visitor mode observing runs in 2007-2008, with corresponding run IDs of 080.A-0427(D)  (0.6 night, Oct 5 2007), 080.A-0427(B) (two half nights 
starting Mar 1, 2008) and 081.A-0103(B) (two half nights starting Aug 24, 2008) on one of the four 8.2m Unit Telescopes of the Very Large Telescope array. A total of 21 MXU (multi-object) masks (6.8$\arcmin$ $\times$ 6.8$\arcmin$ FOV) were observed. Observations were obtained in both the RA14h and RA21h fields, and were designed to maximize the number of extended X-ray sources targeted and their membership. Slits were placed on galaxies with unknown redshifts, prioritizing galaxies close to the X-ray centers and with magnitudes R$<$22, although fainter galaxies were used to fill the masks. A handful of objects with previously determined redshifts were also re-observed to allow for calibration.  

The GRIS300V grism and GG375 filter were used, resulting in an effective wavelength range of $\approx$ 430$-$700 nm.  A slit width of 1$\arcsec$ was used for all objects with a dispersion of 1.68 \AA\ pixel$^{-1}$.  Slit lengths were set to $\geq$ 5$\arcsec$ in order to maximize the total number of objects per mask and obtain many redshifts in the central X-ray source regions where Brightest Group Galaxies (BGGs) are likely to reside. The total integration time per mask ranged from 49 to 77 min.

\subsubsection{FORS2 Data Reduction}
Reduction of the FORS2 data primarily involved a modification of the standard FORS2 pipeline procedure.  The data calibration was performed with version 4.3.5 of the FORS pipeline which performs bias correction, flat-fielding, correction for optical distortions, and wavelength calibration \citep{Appenzeller1998}.  The pipeline also detects and extracts individual object spectra and performs sky subtraction.  However, the standard object detection and sky subtraction pipeline procedures are not ideal for our purposes.  As no co-adding of exposures can be done using the pipeline, chip images from consecutive exposures were co-added using the Image Reduction and Analysis Facility (IRAF\footnote{IRAF is distributed by the National Optical Astronomy Observatories, which are operated by the Association of Universities for Research in Astronomy, Inc., under cooperative agreement with the National Science Foundation.}) \textit{imcombine} tool with cosmic-ray rejection applied before further pipeline reduction steps were applied.  Sky subtraction was done in two ways: first, using a local (on slit) sky subtraction and then using a `global plus median' method.  The latter produces a sky spectrum computed as the median level of all the pixel values of all the CCD spectra in each wavelength bin renormalized after flat-fielding and the other initial processing steps.  This method is the most robust as our small slit size and separation often causes spurious results for local subtraction.  However, for large slits containing faint objects, local subtraction is superior and images processed in this way are used when measuring redshifts for such objects.

In order to determine redshifts, we adapted the ZSPEC software used by the DEEP2 redshift survey \citep{Davis2003,Davis2007} for use with our FORS2 MXU data. In ZSPEC, spectra are first cross-correlated to eigen-templates (stellar, galaxy, and QSO templates) and the ten best-fitted redshifts and $\chi^2$ are provided. The spectra (both 2D and 1D) and their redshift fits were then visually examined in order to determine the correct redshift. Usually the first or second best  $\chi^2$ fit provided is a good fit.   Instances where artifacts from sky line subtraction confuse the fitting, the signal-to-noise of the spectrum is relatively low,  only a single emission line is detected, the spectrum is relatively featureless, or a bad pixel column exists often result in the first or second ranked fits being incorrect.  In these cases, the appropriate solution often appears in a fit with a lower $\chi^2$.  When none of the ten choices is a good match, any spectral features easily identified by eye were used to identify a probable redshift, which could be confirmed by manual cross-correlation. 

Objects without a successful redshift determination were re-evaluated with additional information, such as object magnitude and slit position which can indicate for example that a redshift is unlikely to be obtainable or, in the case of very bright objects a stellar template is preferable, in a final attempt to establish a redshift.  However, most objects for which we could not measure redshifts were very faint or -- in the case of very bright objects -- in an area where sky subtraction was not robust or where extraction was compromised due to slit edge proximity.  At this stage, template fits were possible for the majority of all FORS2 objects ($\sim$970/1270).  Finally, redshifts were assigned a quality flag to reflect the stellar or galactic nature of the object and the confidence of the redshift measurement.  In this analysis we use only the best quality redshifts, excluding objects with ambiguous fits.  In total, 780 high quality galaxy redshifts were obtained. Comparing to CNOC2 and IMACS redshift measurements from duplicate observations, we find a typical error of 100 km/s for our FORS2 redshifts. 

\subsection{IMACS Observations}
We also obtained spectroscopy for both fields using Magellan-IMACS.   The large field of view and close slit placement capability make the IMACS instrument  excellent for observing galaxy groups at intermediate redshift in general and, specifically, its 15$\arcmin$ FOV is an excellent match to that of XMM and thus to our X-ray selected systems follow-up.  Two multi-object masks of the RA14h field  were observed in July 17-18, 2007 on the Baade/Magellan I 6.5m telescope. These were taken with a grism of 200 lines mm$^{-1}$, giving a wavelength range of $\approx$5000 - 9500 \AA\ and a dispersion of 2.0 \AA\ pixel$^{-1}$. A slit width of 1$\arcsec$ was used and the exposure time was two hours for both masks.  For these observations, the WB4800-7800 filter was used.  These observations were made under relatively poor conditions, namely significant moon.  This, and the restricted wavelength range produced by the filter applied, lowered the overall redshift determination success rate for these masks. A further three masks in the RA14h field and two in the RA21h field were obtained in May 18-22, 2009 all with similar setup but without this filter.  

\subsubsection{IMACS Data Reduction}
IMACS data were reduced using the Carnegie Observatories System for MultiObject Spectroscopy (COSMOS\footnote{http://obs.carnegiescience.edu/Code/cosmos}) package. First, overscan regions of the CCDs were used to measure and subtract the bias level. Dome-flat exposures taken during the night were used to flat-field the data. Sky subtraction was performed using the method outlined by \cite{Kelson2003}. Wavelength calibrations were determined from HeNeAr arc exposures. 

Redshifts for the IMACS spectra were determined from cross-correlating the flux-calibrated object spectra with input model templates. The routine adopts SDSS spectral templates for early-type (SDSS template 24) and late-type galaxies (28) as input models and determines the best-fit redshift based on matching absorption and emission line features. The best-fit redshifts returned from the routine was then visually inspected to verify the object's redshift. For some objects, a good template fit was not found by the automatic routine, but spectral features were clearly visible in the galaxy spectrum. In these cases, we performed a manual cross-correlation to determine the redshift. In total, 865 high quality galaxy redshifts, with errors of 140 km/s, were obtained from the IMACS observations.

\subsection{Additional Spectroscopy}

Finally, a single group in the RA14h field (XR14h03) was recently targeted with GMOS-S as part of an ongoing study of galaxy groups within the redshift range 0.85$<$z$<$1 (PI Balogh).   Slit widths were set to 1$\arcsec$ and a R600 grism with OG515 order blocking filter used.  The spectroscopy was obtained in nod \& shuffle mode \citep{Glazebrook2001}, nodding the telescope by $\pm$0.725$\arcsec$ from the centre of the slit, every 60 seconds with a total exposure time of two hours per group target.  All data were reduced in IRAF, using the GEMINI packages with minor modifications.  See \cite{Balogh2011a} for further details of these observations and the data reduction.  In total, 83 high quality galaxy redshifts with errors of 100 km/s were obtained.

\setlength{\tabcolsep}{0.05in} 
\ctable[ caption = {Summary of Supplemental Spectroscopy}, label = tab:spec, maxwidth = 0.49\textwidth, doinside=\small,pos = !htbp, notespar, ]{@{\extracolsep{\fill}} llccccc}{ 
\tnote[*]{4800-7800 for three of the  RA14h field masks\\}
\tnote[]{\textrm{\scriptsize{Column description: Instrument (column 1), wavelength range (2), field of view (3), CNOC2 patch (4), number of masks (5), number of spectra (6), total number of redshifts (7). Note that for IMACS, the number of redshifts excludes stars. }}}
}{  
\hline \hline
 & $\lambda$ range  & FOV & field & N$_{\textrm{\scriptsize{masks}}}$ & N$_{\textrm{\scriptsize{spec}}}$ & N$_{\textrm{\scriptsize{z}}}$ \tstrut \\
 & [\AA] & [$\arcmin$] & &  & &                                                                \\
 \FL
 IMACS & 5000-9500\tmark[*] & 15.5$^\square$ & RA14h & 5 & 1197 & 553     \\
 & & & RA21h &	2 & 551 & 312                                                        \\
 FORS2 & 4300-7000 & 6.8$^\square$ & RA14h & 8 & 520 &  363\\
 & & & RA21h & 13 & 750 & 636                                                             \\
 GMOS & 5000-10000 & 5.5$^\square$  & RA14h & 3 & 125 & 115                     \\
 \hline \hline
}

%SECTION:  NIR OBSERVATIONS AND GALAXY MASSES
\section{NIR Photometry and Stellar Mass}
\label{sec:nirstellmass}

\subsection{NIR Observations}

Details of the near infrared K$_{s}$ observations of the CNOC2 fields from SOFI on the New Technology Telescope (NTT) and Ingrid on the William Herschel Telescope (WHT) can be found in \cite{Balogh2009}. These observations however did not cover much of the area with X-ray coverage (in particular much of the the RA21hr field) and so we have also obtained data with the WIRCam (Wide-field InfraRed Camera) on the Canadian France Hawaii Telescope (CFHT). This data is described by \cite{McGee2011} but a brief description follows here.  Four pointings were made for each of the two fields with each pointing having 33 minutes of exposure time. Each pointing was dithered in a five point pattern to fill in the chip gaps and divided into 80 exposures of 25 seconds each. The resulting coverage area is 30$\arcmin$ $\times$ 30$\arcmin$ per field. These data were subsequently reduced and processed by the Terapix pipeline.

\subsection{Galaxy Stellar Masses}

Stellar masses for our galaxies were computed by template-fitting their spectral energy distributions (SEDs), using available photometry, as in McGee et. al, 2010.  A summary of this stellar mass derivation follows.  The observed photometry,  typically including K, i, r, g, u, GALEX NUV and FUV, was compared to a large grid of model SEDs constructed using the \cite{Bruzual2003} stellar population synthesis code and assuming a Chabrier initial mass function \citep{Chabrier2003}.  This grid of models uniformly samples the allowed parameters of formation time, galaxy metallicity, and the dual component \citet{Charlot2000} dust model and, as in \citet{Salim2007}, the star formation history of a galaxy is assumed to be represented by an exponential model augmented with starbursts.  Model magnitudes at nine redshift bins between 0.25 and 0.6 were derived by convolving these model SEDs with the observed photometric bandpasses . $\chi^{2}$ is minimized while summing over all the models at the redshift of the galaxy and taking the observed uncertainty on each point into account.  Comparison to other estimates of stellar mass shows 1$\sigma$ uncertainties of the order 0.15 dex.

%SECTION:  GROUP REDSHIFTS AND MEMBERSHIP
\section{Measuring Group Properties}
\label{sec:gzmem}

The global properties of our X-ray and optically selected groups are presented in Tables \ref{tab:Xgrp_props1}, \ref{tab:Xgrp_props2},  \ref{tab:Optgrp_props1}, and \ref{tab:Optgrp_props2}.  Table \ref{tab:Xgrp_props1} lists the group identification number for the X-ray system (column 1); IAU name (2); R.A. and Decl. of the center of the extended X-ray emission for Equinox J2000.0 (3 \& 4); spectroscopic redshift (5); group redshift quality (6); the total flux in the 0.5$-$2 keV band (7); and significance of the X-ray flux (8). Table \ref{tab:Xgrp_props2} lists the lists the group identification number for the X-ray system (column 1); number of member galaxies within 1\,Mpc (2); the radius in arcseconds of 1\,Mpc (3); the velocity dispersion within a 1\,Mpc cut (4); spectroscopic completeness (to R$<$22) within a 1\,Mpc cut  (5), dynamical complexity within a 1\,Mpc cut  (6); the number of member galaxies within a $\sigma$ derived r$_{\textrm{\tiny200}}$ (r$_{\textrm{\tiny200,}\textrm{\scriptsize{$\sigma$}}}$) (7); the radius, r$_{\textrm{\tiny200,}\textrm{\scriptsize{$\sigma$}}}$, in arcseconds (8); the velocity dispersion within r$_{\textrm{\tiny200,}\textrm{\scriptsize{$\sigma$}}}$ (9); spectroscopic completeness (to R$<$22) within r$_{\textrm{\tiny200,}\textrm{\scriptsize{$\sigma$}}}$ (10), dynamical complexity within r$_{\textrm{\tiny200,}\textrm{\scriptsize{$\sigma$}}}$ (11);  the number of member galaxies within an X-ray derived r$_{\textrm{\tiny200}}$ (12); the radius, r$_{\textrm{\tiny200,X}}$, in arcseconds (13); velocity dispersion within r$_{\textrm{\tiny200,X}}$ (14); spectroscopic completeness (to R$<$22) within r$_{\textrm{\tiny200,X}}$ (15), and the dynamical complexity within r$_{\textrm{\tiny200,X}}$ (16).  Tables \ref{tab:Optgrp_props1} and  \ref{tab:Optgrp_props2}  tabulate the optical groups and are similarly structured, classifying them according to the group number from the optically selected group catalog of \citeauthor{Carlberg2001} and with an additional column in Tab.\,\ref{tab:Optgrp_props1} (column 5) listing the X-ray group ID where there is a confident match.
  
In total, our sample contains 39 high quality X-ray and 38 optical systems.  Note that IAU names for several groups in the RA21h field have changed  since the publication of Paper I as a result of improved centers due to the addition of XMM data in that field. Two of our high quality systems have low ($<1$) X-ray significance.  The original significance of these systems was sufficient to occasion targeting for follow-up spectroscopy but was subsequently lessened with the addition of X-ray data and modifications to the X-ray reductions.  As follow-up spectroscopy yielded groups with secure redshifts, we choose to include them in our sample, but do not include their X-ray derived properties in our analysis.  12 of our optically selected systems have significant X-ray emission when a fixed aperture is placed at the optical center.

\subsection{Matched X-ray - Optical Systems}
Examining the redshifts and proximity on the sky of groups in both samples, nine of our optically selected systems are associable with X-ray systems but five of these have an X-ray significance $<$ 2 using the optical center and aperture.  Several of our optically selected systems do have significant X-ray emission but are not readily matched to X-ray systems. Recall that the latter require a  $\geq4 \sigma $ detection on the wavelet images in order to be identified.  It is important to note that this is projected X-ray emission and thus absolute certainty in assigning X-ray emission to an optically detected system is not possible. The following sections describe the assignment of redshift and members to all systems as well as the calculation of dynamical mass.

\subsection{X-ray Selected Group Redshifts}
\label{sec:xz}

Initial group redshifts for X-ray systems were established by examining objects within and around the immediate vicinity of the contours defining the extended X-ray sources.  An obvious clustering of galaxies in both redshift and projected spatial coordinates was often obvious and in some cases a prominent galaxy, possibly the BGG, exists near the X-ray center.  Each group was assigned a redshift and a corresponding quality flag reflecting its plausibility.  These quality flags are not related to those for individual galaxy redshifts and range from 1-3, from highest to lowest quality.   A quality of 1 reflects complete confidence in the redshift assignment and significant X-ray emission while a quality of 2 indicates a fairly confident redshift but low significance X-ray emission, weak multiple clustering, or a highly incomplete region within the group (either an extended gap in the survey area or a few bright objects near the group center without redshifts).  A quality of flag 3 indicates a highly questionable redshift due to the projection of strong multiple clustering (more than 1 redshift at which galaxies are grouped), very low significance X-ray emission, or a dearth of objects with redshifts in the area.   We add 0.5 to the quality flag of any group in regions of `confused' emission: where X-ray contours overlap. This does $\it{not}$ indicate a lesser certainty in the redshift / overall quality of the group.  In this paper we will often refer to subsets of X-ray systems as quality 1 and as quality 1 \& 2, ignoring the 0.5 flag.  This flag, and information about a system's dynamical complexity (see  \S \ref{sec:substructure}), allows us to explore underlying reasons for outliers from our relations.

In cases of more than one plausible group redshift, we apply our membership finding algorithm with all possible solutions to find the most self-consistent solution.  Recall that we are always examining projected X-ray emission and that, as in assigning X-ray emission to previously defined optical systems, assigning a group redshift and galaxy members to X-ray emission cannot be done with 100\% confidence.  The difficulty can be illustrated in part by the existence of such a large number of quality 3 groups in our sample many of which are classified as such due to the presence of galaxy clustering at different redshifts.

%PLOT - LX-Z
\begin{figure}[!htb]
\begin{minipage}[t]{1.\linewidth}
      %\vspace{0pt}
      \includegraphics[scale=1, trim = 0 5 15 15,clip]{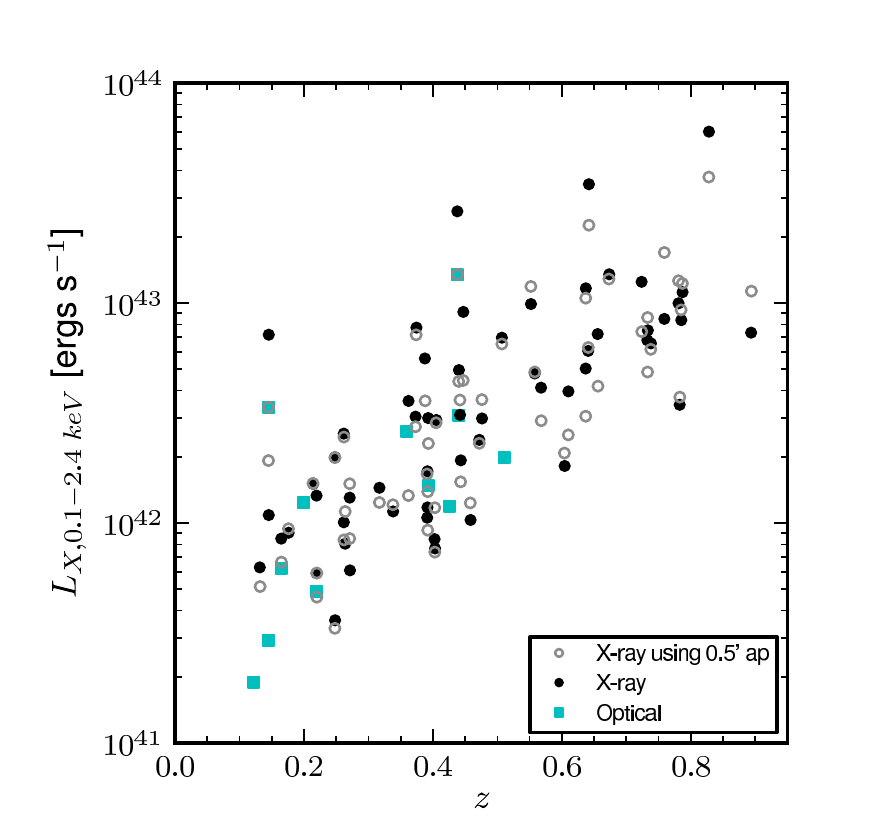}
       \caption{X-ray luminosity as a function of redshift for all X-ray (black circles) and optical (cyan squares) systems with X-ray significance $\ge$ 1 and 2 respectively. Note that 2$\sigma$ upper limits on X-ray luminosity for optical systems with low significance are not shown here.  For X-ray systems, the X-ray luminosity derived using a fixed 0.5$\arcmin$ aperture is shown in grey open circles.}
     \label{fig:fig2_lxz}
\end{minipage} 
%PLOT - SIG-Z
\begin{minipage}[t]{1.\linewidth}
      \includegraphics[scale=1, trim = 0 5 15 10,clip]{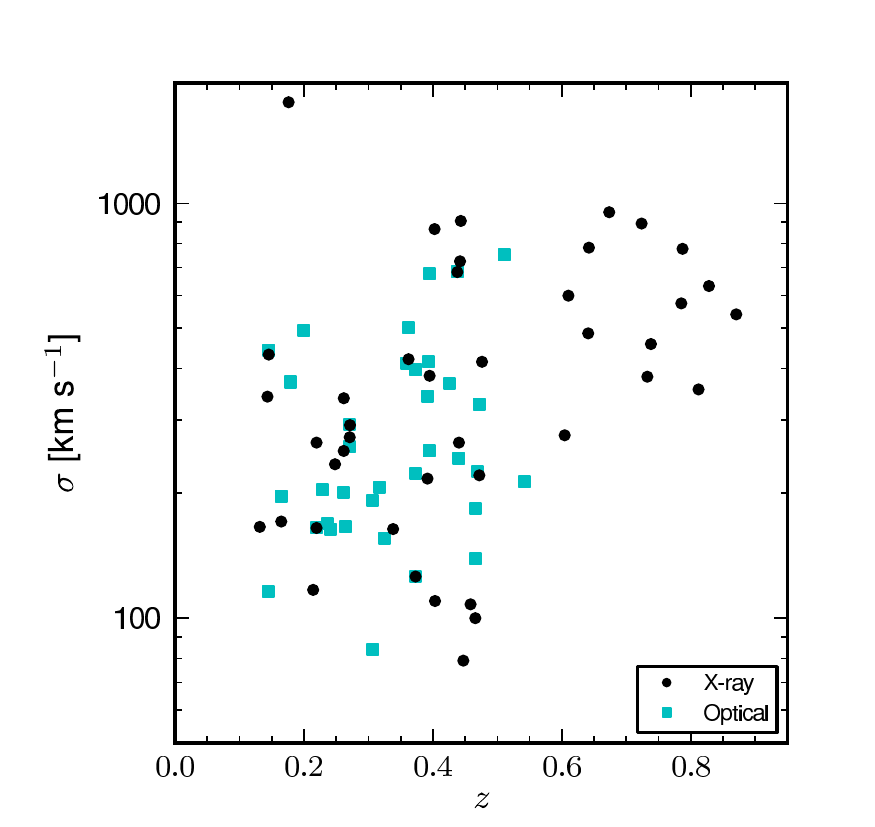}
       \caption{Velocity dispersion as a function of redshift given a $\sigma$ based r$_{\textrm{\tiny200}}$ radial cut (see \S \ref{subsec:rcuts} for details) for all X-ray and optical systems.  Note that upper limits on velocity dispersion are not shown here.}
     \label{fig:fig3_sigz}
\end{minipage}
\end{figure}

\subsection{Group Membership}
\label{sec:mem}
Details of our membership finder algorithm are found in \cite{Wilman2005a} but a basic description, with changes made for the current analysis, follows.  Beginning with the X-ray center and redshift for each group, an initial velocity dispersion of 500 km s$^{-1}$ is assumed and a maximum redshift offset $\delta(z)_{max}$ calculated to clip members at 2 $\times$ the velocity dispersion.  This is then converted into a spatial offset $\delta(\theta)_{max}$ which is within 1/10 the equivalent distance to $\delta(z)_{max}$ in the line-of-sight direction, and group members are selected by applying these redshift and spatial limits as follows:
\begin{equation}
\delta(r)_{max} = \frac{1}{10} \frac{\delta(z)_{max}}{h^{-1}_{75} Mpc}
\end{equation}

\begin{equation}
\delta(\theta)_{max} = 206265\arcsec \frac{\delta(r)_{max}}{h^{-1}_{75} Mpc} \Biggl(\frac{D_{\theta}}{h^{-1}_{75} Mpc}\Biggr)^{-1}
\label{eq:dtheta}
\end{equation}
where 10 is the aspect ratio and the angular diameter distance D$_{\theta}$ is a function of redshift.  Note to tune this offset limit to our X-ray selected group sample, allowing for distinction between adjacent systems while still obtaining stable membership solutions, we tighten it from the value of five used in \citet{Wilman2005a}, choosing instead the aspect ratio of 10.  This aspect ratio is applied to all systems, including those that are optically selected.

In order to obtain an accurate estimate for groups which have relatively few members,  the galaxies within this radius are ordered sequentially by redshift and the observed velocity dispersion, $\sigma(v)_{obs}$, is then calculated using the Gapper algorithm \citep[Eq. 3]{Beers1990} as follows:
\begin{equation}
\sigma(v)_{obs}= 1.135 c \Biggl(\frac{\sqrt{\pi}}{n(n-1)}    \displaystyle\sum_{i=1}^{n-1}{w_{i}g_{i}}       \Biggr)
\end{equation}
where $w_{i}=i(n-1)$ and  $g_{i}=z_{i+1}-z_{i}$ and the 1.135 multiplicative factor corrects for the 2$\sigma$ clipping of a Gaussian velocity distribution.  

This value is then shifted to a rest-frame velocity dispersion and, finally, the intrinsic velocity dispersion $\sigma_{\textrm{\tiny{intr}}}$ is calculated by subtracting the errors from the redshift measurements in quadrature.  The mean redshift of the members and new velocity dispersion is then used to recompute the redshift and spatial offsets and the entire process is repeated until a stable membership solution is attained.  In cases where combined errors from the redshifts measurements are larger than the rest-frame velocity dispersion of the group, we place a 1$\sigma$ upper limit on the intrinsic velocity dispersion using Monte-Carlo simulations. For all other groups, errors on the group velocity dispersion are calculated using the Jackknife technique \citep{Efron1982}.  Note that the assumption of symmetric errors in this case can result in an error measurement larger than the velocity dispersion itself. Fig.\,\ref{fig:fig3_sigz} shows the velocity dispersions for all systems as a function of redshift.

In the rare cases where the algorithm oscillates infinitely between two membership solutions, we choose the solution with more members.  Note that in such cases it is possible that a few member galaxies lay outside the final quoted $\delta(\theta)_{max}$ since this quantity is calculated from the final velocity dispersion of the group.  To evaluate the results of the membership assignment, especially in cases with more than one possible group redshift, we examine both the imaging (X-ray and optical) and velocity distribution of the group members.  

\subsection{Group Centers}

\begin{figure}[!htb]
      \includegraphics[scale=.97,trim = 16 0 10 10,clip]{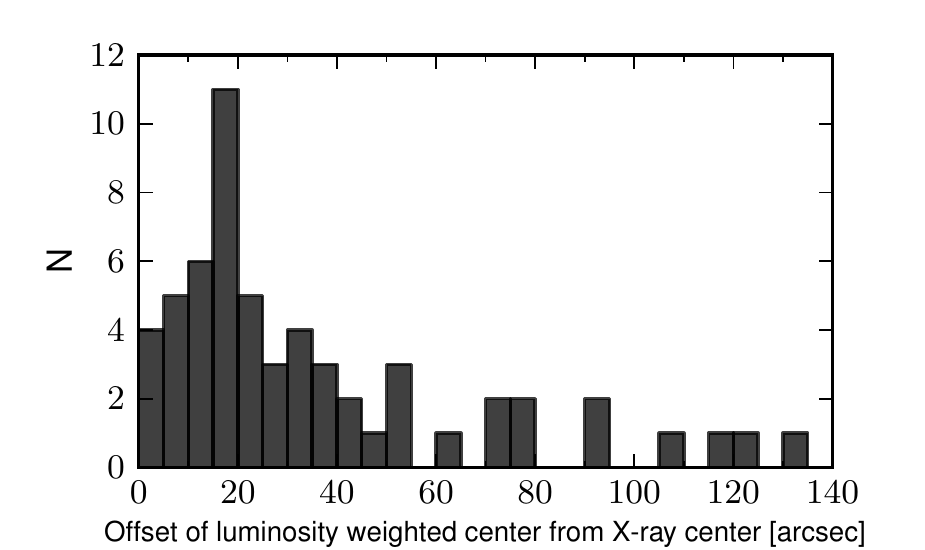}
       \caption{Histogram of offset between X-ray and luminosity weighted centers for all X-ray systems having significance $\ge$1 without a radial cut applied.}
     \label{fig:fig4_cencomp}
\end{figure}

The process of assigning group membership is applied to both the X-ray and optically selected groups, using previously defined optical group centers for the latter.  It is run twice, allowing for R-band luminosity-weighted recentering of the group in the second instance. In Fig.\,\ref{fig:fig4_cencomp} we compare the X-ray and luminosity weighted centers for our X-ray groups.  For the majority of systems, the center shifts $\leq$18$\arcsec$ when luminosity re-centering is applied.  Group membership and overall properties change very little using the luminosity weighted center (see e.g. \S \ref{sec:lxsig}) and we choose to adopt the X-ray centers for these systems in all subsequent analysis.

%PLOT - GROUP CENTER COMPARISON

\subsection{Radial Cuts}
\label{subsec:rcuts}

Once the best redshift and membership is determined, we apply three different radial cuts to our groups: a constant cut of 1\,Mpc and two r$_{\textrm{\tiny200}}$ cuts defining the radius at which the density of the system is 200 times the critical density.  Using the velocity dispersion ($\sigma_{intr}$) and the definition of r$_{\textrm{\tiny200}}$ as in \cite{Carlberg1997}, we define a velocity dispersion based r$_{\textrm{\tiny200}}$ as follows:  
\begin{equation}
\label{r200}
r_{\textrm{\tiny200,}\textrm{\footnotesize{$\sigma$}}}= \frac{\sigma_{\textrm{\tiny{intr}}}\sqrt{3}}{10 H(z)} .
\end{equation}
The second r$_{\textrm{\tiny200}}$ cut is X-ray based and is discussed in \S \ref{sec:xmeas}.  Using each of these radial cuts, the membership is redefined and a final $\sigma$ computed.  In the case of a velocity dispersion based r$_{\textrm{\tiny200}}$, the algorithm is allowed to iterate until a stable solution is found.   Although these cuts often result in decreased membership, if the radial cut is larger than $\delta(\theta)_{max}$, as is often the case for the 1\,Mpc cut, the membership of the group may increase.

For those groups that are adjacent both in position on the sky and in redshift space (e.g. XR21h06 and XR21h07), members may be shared across groups.  As the original membership algorithm does $\it{not}$ allow members to be in multiple groups and instead will merge such systems, discarding one group entirely, the discarded system in these adjacent groups will also have zero members given a r$_{\textrm{\tiny200,}\textrm{\scriptsize{$\sigma$}}}$ radial cut. Distinct groups may also be entirely stripped of members given a r$_{\textrm{\tiny200,}\textrm{\scriptsize{$\sigma$}}}$ if the initial velocity dispersion and subsequent iteration sufficiently reduces the redshift and spatial limits.  This happens most commonly when group members are rather dispersed in projected position and have very similar velocities, causing the computed $\sigma$ and r$_{\textrm{\tiny200,}\textrm{\scriptsize{$\sigma$}}}$ to be low and many, or all, of these members to be discarded.  

As group membership can change significantly given different definitions of group radius, the mean redshift of members may also change.  For the vast majority of systems, the redshift varies little, being stable at least to the third decimal place, and approaches $\sim$0.003 only in the most extreme case.  As the redshift of the group does not affect most quantities subsequently derived for the group, and those that are redshift-dependent are not significantly affected by differences at the level observed, we choose to apply a single redshift in all cases.  X-ray properties (i.e. luminosity and mass) were calculated using the redshift of the group derived with the initial cut, as defined by Eq.\,\ref{eq:dtheta}, and this redshift is typically in very good agreement with those from the three radial cuts.

%SECTION:  MASSES
\section{Group Mass Estimates}
\label{sec:masses}

Table \ref{tab:Xgrp_masses} includes all three mass estimates (X-ray, dynamical, and stellar) for our X-ray selected systems, listing the group identification number for the X-ray system (column 1); rest-frame luminosity in the 0.1$-$2.4 keV band (2); estimates of a total mass, using X-ray luminosity as a mass proxy and a calibration of \cite{Leauthaud2010} (3); group stellar mass calculated using 1\,Mpc, r$_{\textrm{\tiny200,}\textrm{\scriptsize{$\sigma$}}}$, and r$_{\textrm{\tiny{200,X}}}$ radial cuts (4, 5, \& 6); and the dynamical (virial) mass for 1\,Mpc, r$_{\textrm{\tiny200,}\textrm{\scriptsize{$\sigma$}}}$, and r$_{\textrm{\tiny{200,X}}}$ radial cuts (7, 8, \& 9). Table \ref{tab:Optgrp_masses} lists similar quantities for optically selected systems. In cases where there is no significant X-ray detection, we use the upper limit on r$_{\textrm{\tiny{200,X}}}$ in our mass estimates.  Stellar masses are then less than or equal to the derived measurement.  Dynamical masses however may be accurate but could also be under or overestimates as lesser radius could act to increase or decrease $\sigma$ in this case.

\subsection{Dynamical Mass}
We estimate dynamical masses, M$_{dynamical}$ or M$_{dyn}$, for our groups from the velocity dispersion and radius as in \cite{Balogh2006} and \cite{Carlberg1999}: 
\begin{equation}
M_{dyn}=3 \sigma^{2} r_{\textrm{\tiny200}} / G .  
\end{equation}
Note that the factor of three in this equation reflects the assumption of isotropic orbits and an isothermal potential, but is only weakly dependent on those assumptions \citep{Lokas2001}.  We calculate dynamical masses for groups having a minimum of three members. In cases where the velocity dispersion is an upper limit, dynamical masses are also treated as upper limits.  When calculating errors in dynamical mass, no estimation of error in r$_{\textrm{\tiny200}}$ is included.  Large errors in velocity dispersion, which may result from the assumption of symmetric errors made via the Jackknife technique, can result in dynamical mass errors larger than the measure itself.  

\subsection{X-ray Mass}
X-ray masses are estimated using the z$\sim$0.25 relation from \cite{Leauthaud2010}.  Standard evolution of the scaling relations, $M_{\textrm{\tiny200}}E_{z}=f(L_{\textrm{x}} E_{z}^{-1})$ where $E_{z}=(\Omega_M(1+z)^3 +\Omega_\lambda)^{1/2}$, is assumed and these relations verified using a weak lensing calibration of X-ray groups in the COSMOS survey \citep{Leauthaud2010}.  In order to use this calibration, a `concordance' cosmology with H$_0 =$ 72 km s$^{-1}$ Mpc$^{-1}$, $\Omega_M =$ 0.25, and  $\Omega_\Lambda = 0.75$ is applied.  X-ray masses quoted for X-ray and optically selected groups with low ($<$1 and $<$2 respectively) X-ray significance are 2$\sigma$ upper limits.

\subsection{Stellar Mass} 
%\label{subsec:stellmass} 

In order to derive accurate total stellar masses for our groups, we must correct for incompleteness.  The first major contribution to this incompleteness is the lack of spectra for \textit{all} objects in our fields.  To correct for this, we  compute the fraction of objects with redshifts  for each group within its radial cut as a function of R-band (used for spectroscopic selection) magnitude f$_{z}(R)$.  We apply a small correction to this fraction to account for the fact that a small percentage of these objects are likely to be stars.  This minor correction is itself a function of the R-band magnitude and star/galaxy classification.  We then calculate the fraction of members, again as a function of R-band magnitude, by computing the number of known members and dividing this total by the fraction of galaxies having redshifts f$_{mem}(R)$=N$_{mem}(R)$/N$_{z}(R)$.  Finally, the galaxy masses are weighted to correct for this incompleteness as a function of R-band magnitude: weight$_{mem}(R,M_{stellar}) = 1 + (1 - f_{z}(R))/ f_{z}(R) \times f_{mem}(R)$.

\begin{figure}
\includegraphics[scale=.95,trim = 25 10 0 10,clip]{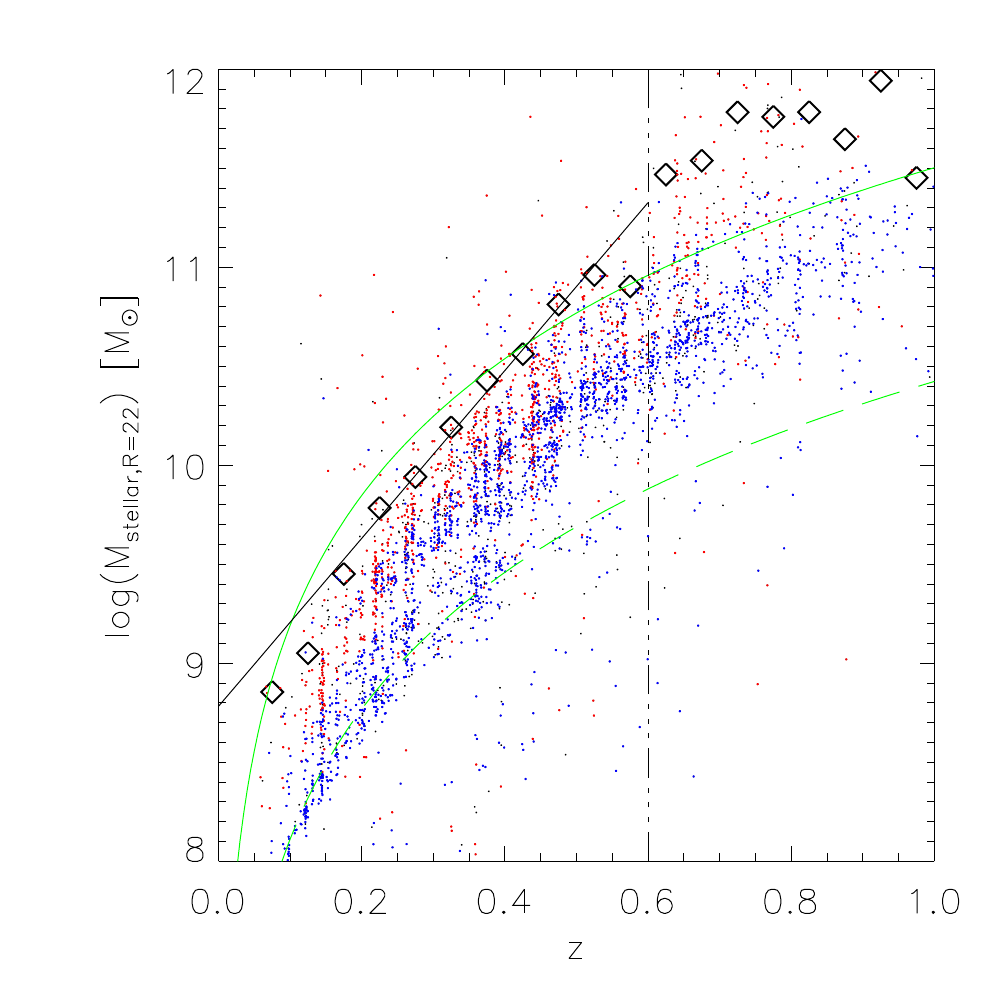}
\caption{Stellar mass limit as a function of redshift. Red and blue dots, respectively, indicate red and blue galaxies, based on rest-frame U--R colors.  Open black diamonds indicate the 90th percentile value of the mass estimates for a red galaxy with R=22, and the black solid line is a simple fit to these points for z$<$0.6.  The vertical dot-dash line indicates z=0.6; no further extrapolation to lower mass is performed at z$>$0.6.  Green solid and dashed lines represent a M/L ratio of 12 and 1 respectively.}
\label{fig:fig5_stellarmasscut}
\end{figure}

The second major source of incompleteness results from the magnitude limit of our spectroscopy.  We begin by recalling our overall R-band magnitude limit of 22.  This limit means that low mass, faint galaxies will be missed.  The mass at fixed magnitude is a function of mass-to-light (M/L) ratio.  Therefore, in order to calculate the appropriate stellar mass limit for each group, we find the mass of a high M/L galaxy at the R=22 magnitude limit as a function of redshift.  Fig.\,\ref{fig:fig5_stellarmasscut} shows the limit in stellar mass as a function of redshift. By examining the distribution of rest-frame U--R color as a function of redshift, we define a line separating the blue and red galaxy populations and categorize all galaxies with U--R $> (0.2 \times z) - 1.5$ as red and the rest as blue. In a given redshift bin, we calculate the mass each galaxy would have if it were observed at the magnitude limit of R=22 and with its own mass-to-light ratio in that band: $M_{stellar, R=22}(z)=M_{stellar}(z)\times10^{-0.4(22.-R(z))}$.  Finally, we compute the 90th percentile value of these mass estimates for the red galaxies in each redshift bin (these are the black diamonds in Fig.\,\ref{fig:fig5_stellarmasscut}) and perform a simple linear fit to these values to define $M_{stellar\_lim}$(z), up to a maximum z=0.6 above which this completeness limit becomes unusefully high.  This fit is comparable to what one would obtain from assuming a mass-to-light ratio of 12.  Finally, we calculate a mass cutoff for each group, $M_{cut, group}$ using its redshift.

%PLOT - MASS CUTOFF

In order to extrapolate the mass below the limit at which we are complete, we first take our lowest redshift groups and fit a Schechter function.  Using the parameters from this local Schechter function fit, we then extrapolate the stellar mass of each group below $M_{cut, group}$ down to a constant cutoff $M_{cut}=10^{10} M_{\sun}$.   We find that the parameters for a system with $log(M_{halo})=13.64$ from \citet{Yang2009},  with $\alpha=-1.22$ and $log(M_{*})=11.122$, provide a reasonable fit for our local groups -- these parameters are then used for the extrapolation.  The final, corrected total group stellar mass is then summed down to our constant mass limit of 10$^{10}$ M$_{\sun}$.  For groups at redshifts z$>$0.6, Schechter function based extrapolation is not used and instead the total measured stellar mass of known members is considered a lower limit.   Note that we calculate stellar masses only for groups with three or more members.   

To calculate errors on the stellar mass determinations for our groups we account for the sampling error by bootstrapping the membership allocation above the mass limit at the group redshift, allowing the galaxies to be selected more than once. We also resample the fraction of galaxies with known redshifts which are members f$_{mem}$(R) selecting from a binomial distribution.  In cases where f$_{mem}$= 0 or 1, we choose to binomially resample the fraction of members presuming that the true fraction is different from these extreme values by 0.5 times the resolution ($\langle$f$_{mem}$$\rangle$ = 0.5/N$_{z}$(R) or 1 -- 0.5/N$_{z}$(R)). For groups with z$<$0.6 where M$_{cut, group}$$>$$10^{10} M_{\sun}$, the extrapolation to lower mass (below the group mass limit) introduces additional uncertainty via the choice of the Schechter function parameters.  To quantify this, we resample the correction randomly from a reasonable range [13.05$<log(M_{halo})<$14.58] of parameter solutions from Tab.\,4 in \cite{Yang2009}.  To quantify the systematic errors associated with the individual galaxy stellar mass measurements, we calculate the group stellar masses using the 2.5 and 97.5 percentile masses from the probability distribution of SED fits to our galaxies and find total group stellar masses an average of 0.5 and 2.0 times those found using the median galaxy masses respectively regardless of the total group stellar mass.

%SECTION:  SUBSTRUCTURE
\section{Dynamical Complexity}
\label{sec:substructure}

\subsection{Descriptions of Tests} 

We search for dynamical complexity\,/\,substructure in our groups by applying the Dressler-Shectman (DS; \citealt{Dressler1988}) Test as in \cite{Hou2012}. The DS Test uses both spatial and velocity information in order to identify substructure. A thorough discussion of this test and its application can be found in \citeauthor{Hou2012} but a brief discussion of our methodology follows. We begin with the mean velocity and velocity dispersion ($\bar{v}$,$\sigma$) for each group having n member galaxies. Then for each galaxy $\it{i}$ in the group, we select it and a number of its nearest neighbours, N$_{nn}$, and compute their mean velocity $v^{i}_{local}$ and velocity dispersion $\sigma^{i}_{local}$ . From these we compute 
\begin{equation}
\delta_{i}^{2} = \Biggl( \frac{N_{nn}+1}{\sigma^2} \Biggr) [(\bar v^{i}_{local}-\bar v)^2 + (\sigma^{i}_{local}-\sigma)^2],
\end{equation}
where 1 $\leq\it{i}\leq$ n$_{members}$ and N$_{nn}$=$\sqrt{n_{members}}$. The Dressler-Shectman $\Delta$ statistic is then calculated as follows 
\begin{equation}
\Delta = \sum_{i=1}^N \delta_i,
\end{equation}
where N is the total number of galaxies in the group.

100,000 Monte Carlo models are then run to calibrate the $\Delta$ statistic for each group. Each Monte Carlo model is made by randomly shuffling the velocities among the group galaxies. Then a probability P is defined as the fraction of the total number of Monte Carlo models of the group that have $\Delta$'s larger than the true value of the group. $P \approx 1.0$ means that the group contains no substructure, while $P \approx 0.0$ indicates that the group contains statistically significant substructure.  For a group to be defined as having substructure, we require  $P < 0.01$.

Another method of identifying dynamical complexity within groups is to search for deviations from a Gaussian velocity distribution.  We use the Anderson-Darling (AD) Test to classify velocity distributions as non-Gaussian as in \citet{hou09}, and show that the test is reliable and robust for group-sized systems.  A detailed analysis of the AD test is given in \citet{hou09}, but we give a brief description of the statistic here.  The AD statistic is a goodness-of-fit test that compares the cumulative distribution function (CDF) of ordered data to a model empirical distribution function (EDF), which in our case is a Gaussian EDF.  This comparison is done using the following computing formulae \citep{D'agostino}

\begin{equation}
\centering
A^{2}= -n-\frac{1}{n}\sum_{i=1}^n(2i-1)(\ln\Phi(x_{i}) + \ln(1 - \Phi(x_{n+1-i}))),
\label{ADtest}
\end{equation}
\begin{equation}
\centering
A^{2*} = A^{2}\left(1 + \frac{0.75}{n} + \frac{2.25}{n^{2}}\right)
\label{A2star}
\end{equation}
where $x_{i} \leq x < x_{i+1}$, $\Phi(x_{i})$ is the CDF of the hypothetical underlying distribution.  Probabilities for the AD test are then computed using

\begin{equation}
\centering
\alpha = a\exp(-A^{2*}/b) 
\label{alpha}
\end{equation}
where a = 3.6789468 and b = 0.1749916, and both factors are determined via Monte Carlo methods \citep{nelson98}.  A system is then considered to have a non-Gaussian velocity distribution, and therefore dynamical complexity, if its computed $\alpha$ value is less than 0.01, corresponding to a 99 per cent confidence level.  

Hou et al. find both tests to be reliable for groups with 10 or more members, thus we apply them only to those groups in our samples which meet this criterion.  Tests using mock catalogs indicate that this criterion, combined with the requirement of a probability less than 0.01, results in a false positive rate of 1\% and  5\% for the AD and DS tests respectively.

\setcounter{table}{8}
\setlength{\tabcolsep}{0.04in}
\ctable[caption = {Summary of Dynamical Complexity Test Results},label = tab:substruc,maxwidth=0.49\textwidth,doinside=\small,pos = !htbp, notespar,]{lllllll}{
\tnote[*]{quality 1 \& 2 systems only\\}
\tnote[]{\textrm{\scriptsize{Column description: Number of systems meeting dynamical complexity criterion per total number tested for X-ray and optical systems in each radial cut.}}}\\
}{
\hline \hline
& AD & DS & AD & DS & AD & DS \tstrut \\
& 1\,Mpc & 1\,Mpc & r$_{\textrm{\tiny200,}\textrm{\scriptsize{$\sigma$}}}$ & r$_{\textrm{\tiny200,}\textrm{\scriptsize{$\sigma$}}}$ & r$_{\textrm{\tiny200,X}}$ & r$_{\textrm{\tiny200,X}}$ \\
\FL
X-ray\tmark[*] & 7 of 19 & 4 of 19 & 6 of 14 & 4 of 14 & 1 of 11 & 0 of 11 \\
Optical & 5 of 19 & 3 of 19 & 4 of 12 & 1 of 12 & 3 of 10 & 0 of 10 \tstrut \\
\hline \hline
}

\subsection{Effect of Dynamical Complexity} 

Tab.\,\ref{tab:substruc} summarizes our results for both X-ray and optical groups, giving the number of systems where dynamical complexity was identified per the systems tested for each test and radial cut. In general, we find the least amount of dynamical complexity when we employ an X-ray based r$_{\textrm{\tiny200}}$ cut to our systems.  In fact the DS test fails to find significant substructure for any group with this radial cut applied. When we use the, normally larger, 1\,Mpc and velocity dispersion based radial cuts we find significantly more dynamical complexity.  The latter cut yields the highest fraction of both non-Gaussian (AD) and substructure (DS) groups.

\begin{figure}[htbp]
\begin{minipage}[t]{1.\linewidth}
      \includegraphics[scale=0.465, trim = 25 0 0 0 ]{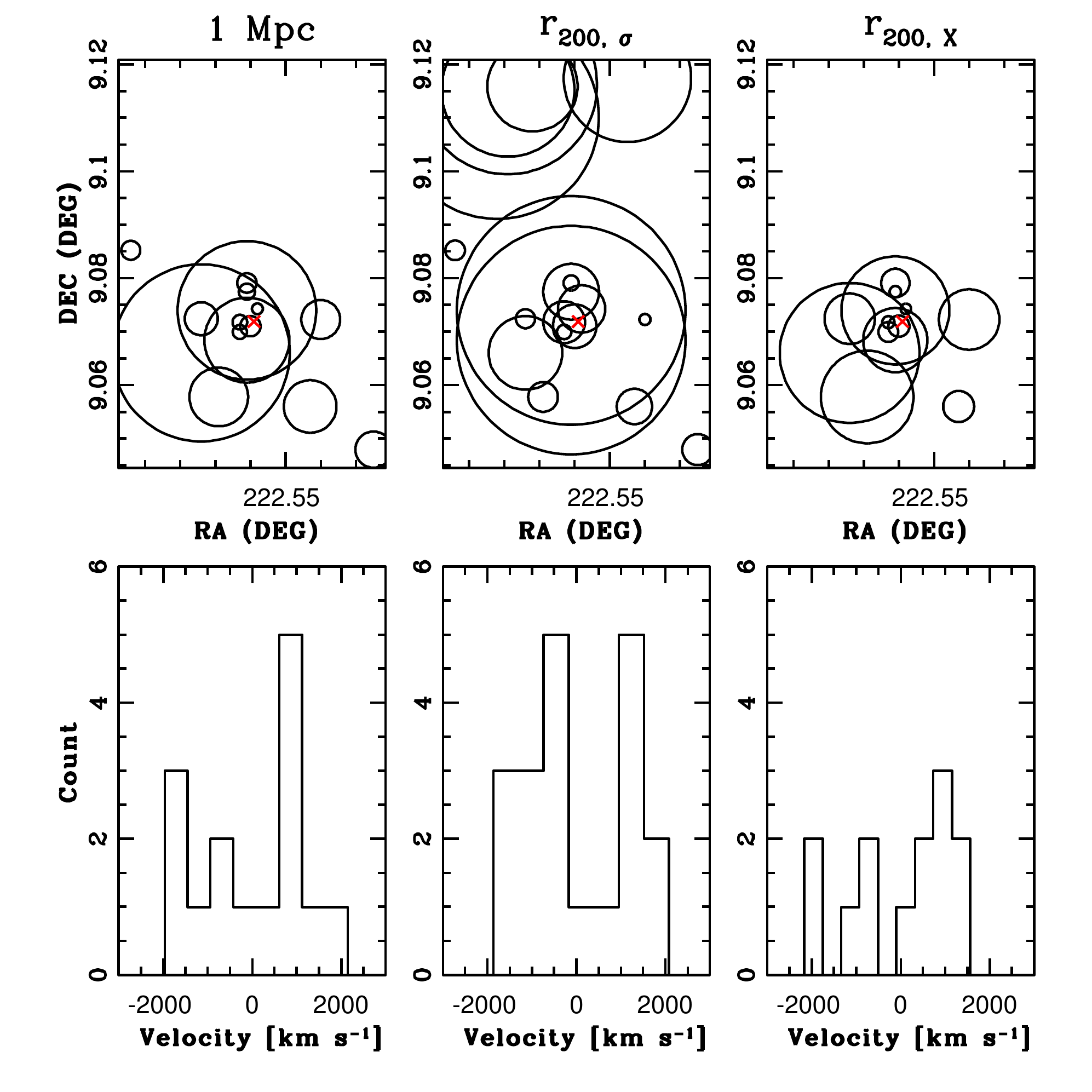}
      \caption{Results of substructure tests for group XR14h09.  Top panel: \cite{Dressler1988} Ôbubble-plotsÕ where the galaxy symbols scale with exp($\delta_{i}$) for 1\,Mpc, r$_{\textrm{\tiny200,}\textrm{\scriptsize{$\sigma$}}}$, and r$_{\textrm{\tiny200,X}}$ radial cuts.  The DS test finds substructure only in the case of an r$_{\textrm{\tiny200,}\textrm{\scriptsize{$\sigma$}}}$ radial cut. Bottom panel: Histogram of the velocity distribution for the same radial cuts as above.  Non-Gaussianity (dynamical complexity) is detected using the AD test at all radial cuts.}
      \label{fig:fig6_sub}
  \end{minipage}
\end{figure}

Fig.\,\ref{fig:fig6_sub} shows the results for both substructure tests for X-ray selected system XR14h09. Substructure is detected in this group at all radial cuts by the AD test and for the r$_{\textrm{\tiny200,}\textrm{\scriptsize{$\sigma$}}}$ cut case according to the DS test. The top panel of Fig.\,\ref{fig:fig6_sub} shows Dressler-Schectman `bubble-plotsÕ for this group for each of the three radial cuts.  In a `bubble-plot', each galaxy in the group is plotted at it's spatial position and is represented by a symbol whose size scales with it's $\delta_{i}$ value.  Larger symbols indicate larger deviations in the local kinematics compared to the global values, and a `local grouping' of galaxies with similarly large symbols may indicate a kinematically distinct system.  The r$_{\textrm{\tiny200,}\textrm{\scriptsize{$\sigma$}}}$ cut plot shows such a congregation at a declination of $\sim$9.115$^{\circ}$.

%PLOT - SUBSTRUCTURE

\begin{figure}[!htbp]
  \begin{minipage}{1.\linewidth}
      \includegraphics[scale=1.,trim=0 5 0 15,clip]{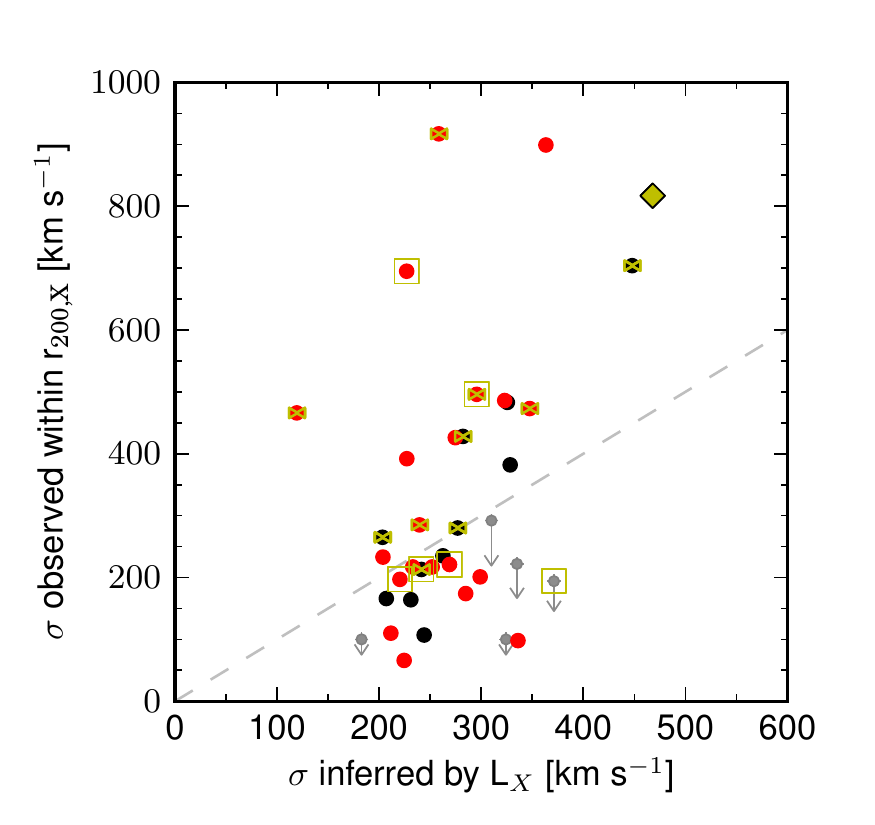}
      \caption{Velocity dispersions for quality 1 \& 2 X-ray selected systems (black and red filled circles respectively) measured within r$_{\textrm{\tiny200,X}}$ and calculated as in Eq.\,\ref{r200} using r$_{\textrm{\tiny200,X}}$. Upper limits are shown in grey.  Yellow bow-ties indicate systems tested for substructure.  Filled yellow diamonds and squares indicate systems with substructure according to AD and DS tests respectively.  All substructure results shown here are for the r$_{\textrm{\tiny200,X}}$ radial cut case.  Open yellow squares show groups in X-ray confused regions.  A 1:1 line is shown in dashed grey.}
      \label{fig:fig7_sigsig}
  \end{minipage}
\end{figure}

The DS test fails to detect substructure for all groups when an X-ray based r$_{\textrm{\tiny200}}$ radial cut applied.  Assuming substructure is preferentially located at the outskirts of groups, this supports the hypothesis that an X-ray based r$_{\textrm{\tiny200}}$ cut is the one most likely to be tracing the virialized core of the system.  Fig.\,\ref{fig:fig7_sigsig} looks more closely at this r$_{\textrm{\tiny200,X}}$ cut, comparing the `real' velocity dispersions \textit{measured} for quality 1 \& 2 X-ray systems within the X-ray defined r$_{\textrm{\tiny200}}$ to the velocity dispersion for the same systems if $\sigma$ were instead \textit{computed} by substituting r$_{\textrm{\tiny200,X}}$ into Eq.\,\ref{r200} and rearranging to get $\sigma$.   The latter results in a much tighter range in velocity dispersions.  Very low velocity dispersions are not possible when inferred from L$_{X}$ due to the X-ray detection limit but measured $\sigma$ can be much larger than that inferred from the X-ray measurements.  This implies that dynamical complexity may inflate velocity dispersion even within an X-ray derived r$_{\textrm{\tiny200}}$ but this is confirmed by the AD test in only a single case.

The differences in $\sigma$ -- and thus dynamical mass -- \textit{measurements} within the different radial cuts are, however, small.   Stellar mass measurements though may be biased since larger radial cuts will always result in larger or equivalent stellar masses.

%PLOT - SIGMA-SIGMA

Although our X-ray imaging may not be deep enough to detect X-ray `substructure' in the majority of our systems, a pair of X-ray detected systems, XR21h06 and XR21h07, does appear to be one clear case of this in our sample.   These groups lie at essentially the same redshift of 0.145 and appear as two separate peaks within an area of overlapping X-ray emission.   The group XR21h07 corresponds to the center of optical group OP21h104.  None of these groups have substructure detected using the AD or DS tests.  This is especially surprising given a 1\,Mpc radial cut as the membership overlaps so thoroughly in this case.  If we relax our criterion to $P \approx 0.06$, XR21h06 would have substructure detected by the DS test in the case of all radial cuts and XR21h07 in the case of an X-ray based r$_{\textrm{\tiny200}}$ cut.  OP21h104 would also have substructure detected via the DS test if the criterion was relaxed and an X-ray based r$_{\textrm{\tiny200}}$ cut employed.  If these groups are merging in the plane of the sky, it is doubtful substructure would be detected by the DS test.  Although the AD and DS tests have been shown to be reliable in most cases, there are certain scenarios in which either/both could fail and one such example is a merger in the plane of the sky.  Since the aforementioned tests essentially look for deviations in the velocity distribution, these types of mergers may not have substructure with significant (or detectable) differences in velocity \citep[e.g.][]{Pinkney1996}.  Further discussion of false negatives for the DS test and the effect of superposition for massive GEEC groups can be found in \cite{Hou2012}.

%SECTION:  Lx-SIGMA
\section{L$_{X}$-$\sigma$ Relation}
\label{sec:lxsig}

\begin{figure*}[!htbp]
\includegraphics[scale=.88 , trim = 0 30 0 35,clip]{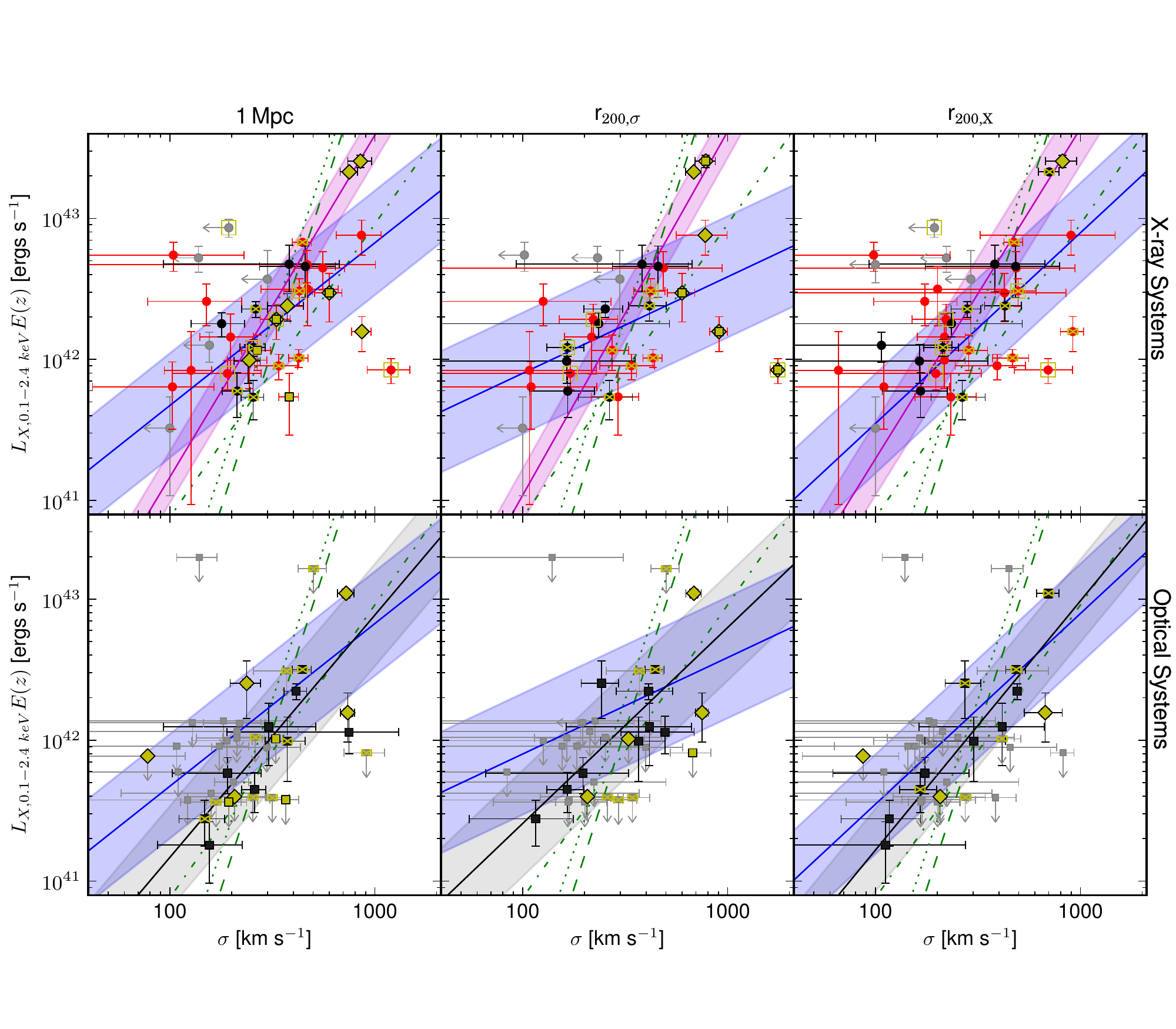}
\caption{L$_{X}$-$\sigma$ relation for quality 1 \& 2 (black and red circles respectively) X-ray selected systems (top) and optical systems (bottom) for all radial cuts. Grey arrows indicate limits. Dashed and dot-dashed green lines show z$\sim$0 sample fits [\citealt{Mulchaey2003}] while the dotted line is a z$\sim$0.25 sample [\citealt{Rykoff2008}].   Bayesian best fits for quality 1 (magenta) and quality 1 \& 2 (blue) X-ray and optical systems (black) are shown with filled region representing the scatter.  Yellow bow-ties show systems tested for substructure.  Filled yellow diamonds and squares indicate systems with substructure according to AD and DS Tests respectively.  Open yellow squares show groups in X-ray confused regions.} 
\label{fig:fig8_lxsig}
\end{figure*}

The X-ray luminosity-velocity dispersion (L$_{X}$-$\sigma$) relation is shown in Fig.\,\ref{fig:fig8_lxsig} for X-ray and optically selected groups with all radial cuts.  In order to define the linear best fit for these relations while accounting for the errors in both L$_{X}$ and $\sigma$, we choose a Bayesian approach as in \cite{Kelly2007}.  Specifically, we use the LINMIX\_ERR IDL code of \cite{Kelly2007} to determine the slope $m$, intercept $c$, and intrinsic scatter $s$ of the relation $\log$(L$_{X}$) $= m \times \log(\sigma) + c + \epsilon$, where $\epsilon$ is a random variable with variance equal to the square of the intrinsic scatter ($s^2$).  The Kelly method allows measurement errors to be treated as independent and log-normal and assumes that the intrinsic scatter in the dependent variable is Gaussian and the intrinsic distribution of the independent variables can be well approximated by a combination of Gaussians.  The publicly available LINMIX\_ERR code constructs Monte Carlo Markov Chains (MCMCs, \citealt{Gilks1996,Christensen2000}) to draw random parameter sets from the probability distributions, the maxima of the distributions of these draws representing the best fit values.  Note the probability distribution can be asymmetric.  We compute 1$\sigma$ uncertainties using 15.9 and 85.1 percentile values of all fit quantities.  Using the resultant slope, intercept, and intrinsic scatter, and their uncertainties, we shall search for increased scatter / outliers from the relation, attempting to tie this to an observable property of the outlying groups.  

\tabletypesize{\scriptsize}
\setlength{\tabcolsep}{0.0275in}
\begin{table}[!htbp]
\caption{ L$_{X}$-$\sigma$ Relation Bayesian Best Fits}
\begin{center}
\begin{tabular*}{1.\textwidth}{l l|c|c|c|}
\cline{3-5}
& & $m$ & $c$ & $s$ \tstrut \\ [1.5pt]
\cline{1-5}
\multirow{3}{*}{X-ray Q=1}
& 1Mpc & 2.5816$\pm^{0.4600}_{0.4435}$ & 35.872$\pm^{1.1489}_{1.1891}$ & 0.1884$\pm^{0.1286}_{0.0916}$  \tstrut \\ [1.5pt] \cline{2-5}
& r$_{\textrm{\tiny{200,}\textrm{\scriptsize{$\sigma$}}}}$ & 2.3432$\pm^{0.5045}_{0.6154}$ & 36.608$\pm^{1.4957}_{1.3824}$ & 0.2592$\pm^{0.1781}_{0.1286}$  \tstrut \\ [1.5pt] \cline{2-5}
& r$_{\textrm{\tiny{200,X}}}$ & 2.4044$\pm^{0.5879}_{0.6071}$ & 36.341$\pm^{1.5298}_{1.5406}$ & 0.2266$\pm^{0.1639}_{0.1126}$  \tstrut \\ [1.5pt]
\cline{1-5}
\multirow{3}{*}{X-ray Q=1 \& 2}
& 1Mpc & 1.1539$\pm^{0.3806}_{0.3793}$ & 39.364$\pm^{0.9873}_{0.9635}$ & 0.3703$\pm^{0.0654}_{0.0533}$  \tstrut \\ [1.5pt] \cline{2-5}
& r$_{\textrm{\tiny{200,}\textrm{\scriptsize{$\sigma$}}}}$ & 0.6844$\pm^{0.3535}_{0.3491}$ & 40.532$\pm^{0.9050}_{0.9149}$ & 0.4303$\pm^{0.0821}_{0.0613}$  \tstrut \\ [1.5pt] \cline{2-5}
& r$_{\textrm{\tiny{200,X}}}$ & 1.3529$\pm^{0.4249}_{0.4650}$ & 38.839$\pm^{1.2093}_{1.0722}$ & 0.3533$\pm^{0.0664}_{0.0548}$  \tstrut \\ [1.5pt] 
\cline{1-5}
\multirow{3}{*}{Optical}
& 1Mpc & 1.7125$\pm^{0.5883}_{0.5902}$ & 37.769$\pm^{1.4878}_{1.5344}$ & 0.3577$\pm^{0.1554}_{0.1034}$  \tstrut \\ [1.5pt] \cline{2-5}
& r$_{\textrm{\tiny{200,}\textrm{\scriptsize{$\sigma$}}}}$ & 1.3628$\pm^{0.6351}_{0.6117}$ & 38.691$\pm^{1.5328}_{1.6341}$ & 0.3779$\pm^{0.1568}_{0.1081}$  \tstrut \\ [1.5pt] \cline{2-5}
& r$_{\textrm{\tiny{200,X}}}$ & 1.7822$\pm^{0.6019}_{0.5350}$ & 37.665$\pm^{1.3516}_{1.5601}$ & 0.2994$\pm^{0.1614}_{0.1119}$  \tstrut \\ [1.5pt]
\cline{1-5}
\end{tabular*}
\label{tab:lxsigfits}
\end{center}
\tnote[]{Column description: Bayesian best fit slope ($m$) and uncertainties (column 1); intercept ($c$) and uncertainties  (2); and intrinsic scatter ($s$) and uncertainties (3) of the relation $\log$(L$_{X}$) $= m \times \log(\sigma) + c + \epsilon$, where $\epsilon$ is a random variable with variance equal to $s^2$.}
\end{table}

For the X-ray groups, we perform this best fit analysis for quality 1, 1 \& 2, and 1, 2, \& 3 groups respectively and for each radial cut excluding those with upper limits on L$_{X}$ and/or $\sigma$.  As expected, the intrinsic scatter in the relation tends to increase with the addition of the poorer quality groups.  Additionally, we examine the effect of luminosity-weighted recentering for X-ray systems by recomputing the membership and velocity dispersion and find little change in $\sigma$ and thus little difference in the L$_{X}$-$\sigma$ relation.  The best fit analysis is also performed for the optical systems for all radial cuts.  For a given radial cut, the L$_{X}$-$\sigma$ best fits for Q=1 \& 2 X-ray and optically selected systems are relatively similar but the relation found for Q=1 X-ray systems significantly steeper.  For a fixed L$_{X}$, the range in $\sigma$ is much larger for Q=2 than for Q=1 X-ray systems. Additionally, the higher $\sigma$ Q=2 groups tend to lie well off the relation and exhibit dynamical complexity. The intrinsic scatter for the optical groups is larger than that found for the quality 1 X-ray groups, regardless of the quality cut applied, but generally comparable or less than that for the quality 1 \& 2 X-ray systems.  Note that most optical systems are excluded from the fitting due to their  L$_{X}$ measurement limits.  We provide the L$_{X}$-$\sigma$ slope, intercept, and intrinsic scatter, and their uncertainties, for the optical and high quality X-ray systems for each of the different radial cuts in Tab.\,\ref{tab:lxsigfits}.

%PLOTS - LX-SIGMA

We plot for comparison the L$_{X}$-$\sigma$ relation derived from two different samples: the \cite{Mulchaey2003} sample of groups at z$\approx$0 and the \cite{Rykoff2008} maxBCG sample of clusters at $0.1$$\le$$z$$\le$$0.3$. We show these comparisons as dashed and dotted lines in Fig.\,\ref{fig:fig8_lxsig} and find no evidence for deviation from the assumed evolution of this relation with redshift. The slope in our relation however clearly depends on the groups selected (i.e. optical, X-ray Q=1, or X-ray systems).   For example, our slope for the highest quality X-ray selected systems is very similar to the 2.56$\pm$0.56 found by Osmond \& Ponman for the GEMS group sample \citep{OsmondPonman2004} while the slopes for our Q=1 \& 2 X-ray systems and optical systems with an X-ray derived r$_{\textrm{\tiny200}}$ are in relatively good agreement with that of \citet[$m=1.7\pm0.4$]{Jeltema2008} though their relation is calculated within r$_{\textrm{\tiny{500,X}}}$ and the groups used are on average more massive than ours.  The group selection then drives a range of slopes which are consistent with both of these results.  In general, our work supports the findings that the L$_{X}$-$\sigma$ relation for groups is shallower than that for clusters \citep{Mahdavi2001,XueWu2000} where the relation has been found to agree well with the bolometric X-ray luminosity $\propto\sigma^{4}$ predicted by self-similar evolution \citep[e.g.][]{Horner2001,Zhang2011}.

Note that several of the most significant outliers at high mass in these relations show substructure (marked as filled yellow diamonds and squares).  Given a $\sigma$ based r$_{\textrm{\tiny200}}$ radial cut, all systems with a velocity dispersion greater than 500 km s$^{-1}$ show dynamical complexity implying that these high values may be overestimated.  This translates to a dynamical mass of $\sim$10$^{14.1}$ M$_{\sun}$. 

High quality X-ray groups in X-ray confused regions (quality 1.5 or 2.5, shown as open yellow squares in Fig.\,\ref{fig:fig8_lxsig}) do not seem to be preferentially high in X-ray luminosity.  For both X-ray (quality 1 \& 2) and optical systems, the use of a $\sigma$ derived r$_{\textrm{\tiny200}}$ cut results in the largest scatter in the L$_{X}$-$\sigma$ relation and looks to be biased towards giving higher dispersions for dynamically complex systems.  The X-ray derived r$_{\textrm{\tiny200}}$ cut provides a relatively tight correlation even for optically selected systems.  A 1\,Mpc cut produces similarly tight fits for the good quality X-ray systems but is less well constrained than the X-ray radial cut for optical systems.  This constant cut can extend beyond a physical r$_{\textrm{\tiny200}}$ or lie within it and is biased large (small) for low (high) halo masses.

As velocity dispersions for systems having few members are less reliable \citep[e.g.][]{ZabludoffMulchaey1998,GirardiMezzetti2001}, we performed additional Bayesian fitting to the L$_{X}$-$\sigma$ relation further dividing our subsamples into those with N$_{mem}<10$ and those with N$_{mem}\ge10$.  Note that the latter subsample is comprised of all groups where dynamical complexity could be evaluated which are thus marked by yellow bow-ties or filled squares or diamonds in Figures \ref{fig:fig7_sigsig} and \ref{fig:fig8_lxsig}.  It is clear from Fig.\,\ref{fig:fig8_lxsig} that, for X-ray selected groups, the low N$_{mem}$ systems  tend to have lower dispersion than the high N$_{mem}$ groups at fixed L$_{X}$ regardless of the radial cut applied. This indicates either that the dispersion and number of members correlates better with the group mass than L$_{X}$ or that the dispersion is typically estimated lower with fewer members.  Nonetheless, the overall fits are within the range of those found for the total population.  As systems with few members also have large measurement errors  in their velocity dispersions, they are consistent with a wide range of relations, while the high N$_{mem}$ groups have smaller errors and thus import more stringent constraints on the best fit relation.  In this light, it is not surprising that similar results are produced.

\tabletypesize{\scriptsize}
\setlength{\tabcolsep}{0.05in}
\begin{table*}[!hbtp]
\caption{ L$_{\textrm{x}}$-$\sigma$ Relation Bayesian Best Fits with Groups Subdivided at N$_{mem}=10$}
\begin{center}
\begin{tabular*}{0.8\textwidth}{l r|c|c|c|c|c|c|}
\cline{3-8}
\multicolumn{2}{c|}{}&\multicolumn{2}{c|}{$m$}&\multicolumn{2}{c|}{$c$}&\multicolumn{2}{c|}{$s$} \\
\cline{3-8}
& & N$_{mem}\ge10$ & N$_{mem}<10$ & N$_{mem}\ge10$ & N$_{mem}<10$ & N$_{mem}\ge10$ & N$_{mem}<10$ \tstrut \\ [1.5pt]
\cline{1-8}
\multirow{3}{*}{X-ray Q=1}
& 1Mpc & 2.7432$\pm^{0.4646}_{0.4385}$ & \nodata & 35.410$\pm^{1.1414}_{1.1809}$ & \nodata & 0.1970$\pm^{0.1726}_{0.0971}$ & \nodata \tstrut \\ [1.5pt]   \cline{2-8}
& r$_{\textrm{\tiny{200,}\textrm{\scriptsize{$\sigma$}}}}$ & 2.6415$\pm^{3.1382}_{1.9239}$ & 1.3271$\pm^{2.6441}_{1.6924}$ & 35.715$\pm^{5.0870}_{8.6822}$ & 39.005$\pm^{4.1871}_{6.3108}$ & 0.9425$\pm^{3.2969}_{0.5930}$ & 0.4254$\pm^{0.4463}_{0.2154}$  \tstrut \\ [1.5pt] \cline{2-8}
& r$_{\textrm{\tiny{200,X}}}$ & 2.7481$\pm^{1.2929}_{0.9573}$ & 1.0101$\pm^{2.3692}_{1.6856}$ & 35.446$\pm^{2.5106}_{3.4051}$ & 39.823$\pm^{4.0695}_{5.5660}$ & 0.4255$\pm^{0.6163}_{0.2428}$ & 0.4703$\pm^{0.5544}_{0.2401}$  \tstrut \\ [1.5pt] 
\cline{1-8}
\multirow{3}{*}{X-ray Q=1 \& 2}
& 1Mpc & 1.8992$\pm^{0.5452}_{0.5434}$ & 0.4102$\pm^{0.4228}_{0.4068}$ & 37.382$\pm^{1.4192}_{1.4129}$ & 41.365$\pm^{1.0160}_{1.0802}$ & 0.3753$\pm^{0.0981}_{0.0763}$ & 0.3720$\pm^{0.1108}_{0.0828}$  \tstrut \\ [1.5pt] \cline{2-8}
& r$_{\textrm{\tiny{200,}\textrm{\scriptsize{$\sigma$}}}}$ & 0.6790$\pm^{0.6856}_{0.6583}$ & 1.3627$\pm^{2.8705}_{5.1712}$ & 40.551$\pm^{1.7749}_{1.8619}$ & 38.960$\pm^{12.436}_{6.8315}$ & 0.5935$\pm^{0.1781}_{0.1215}$ & 0.2772$\pm^{0.1314}_{0.1225}$  \tstrut \\ [1.5pt] \cline{2-8}
& r$_{\textrm{\tiny{200,X}}}$ & 1.6872$\pm^{0.9013}_{0.8285}$ & 0.7869$\pm^{1.0804}_{1.0626}$ & 37.982$\pm^{2.2048}_{2.3799}$ & 40.305$\pm^{2.5923}_{2.6528}$ & 0.5120$\pm^{0.1885}_{0.1263}$ & 0.3212$\pm^{0.0921}_{0.0772}$  \tstrut \\ [1.5pt]  
\cline{1-8}
\multirow{3}{*}{Optical}
& 1Mpc & 1.3423$\pm^{1.6965}_{1.7528}$ & 1.6129$\pm^{1.5271}_{1.2340}$ & 38.801$\pm^{4.6197}_{4.4914}$ & 37.955$\pm^{2.9675}_{3.8198}$ & 0.7150$\pm^{0.7264}_{0.2929}$ & 0.4429$\pm^{0.5366}_{0.2305}$  \tstrut \\ [1.5pt] \cline{2-8}
& r$_{\textrm{\tiny{200,}\textrm{\scriptsize{$\sigma$}}}}$ & 1.1933$\pm^{1.4007}_{1.4801}$ & 1.1905$\pm^{3.6719}_{2.8864}$ & 39.256$\pm^{3.8189}_{3.6676}$ & 38.973$\pm^{7.0081}_{9.4757}$ & 0.6992$\pm^{0.7714}_{0.2927}$ & 0.6732$\pm^{1.8007}_{0.4107}$  \tstrut \\ [1.5pt] \cline{2-8}
& r$_{\textrm{\tiny{200,X}}}$ & 1.6430$\pm^{1.1719}_{1.1730}$ & 1.2470$\pm^{1.7510}_{1.8734}$ & 38.067$\pm^{3.0628}_{3.0917}$ & 38.922$\pm^{4.3277}_{4.4495}$ & 0.5334$\pm^{0.5298}_{0.2385}$ & 0.7305$\pm^{1.5954}_{0.4627}$  \tstrut \\ [1.5pt] 
\cline{1-8}
\end{tabular*}
\end{center}
\label{tab:lxsigfits_10mem}
\tnote[]{\scriptsize Column description: Bayesian best fit slope ($m$) and uncertainties (columns 1 \& 2); intercept ($c$) and uncertainties (3 \& 4); and intrinsic scatter ($s$) and uncertainties (5 \& 6) of the relation $\log$(L$_{X}$) $= m \times \log(\sigma) + c + \epsilon$, where $\epsilon$ is a random variable with variance equal to $s^2$.  The first column of each quantity is for groups with at least ten members while the second only includes those with less than this amount.  Note that in the case of the quality 1 X-ray groups with a 1\,Mpc radial cut, there is an insufficient number of groups with less than ten members to perform robust fitting.}
\end{table*}

 %SECTION: TOTAL MASS MEASUREMENTS
 \section{Total Mass Measurements}
 \label{sec:totmass}

\begin{figure}[!htbp]
\begin{minipage}{1.\linewidth}
\includegraphics[scale=1, trim = 10 7 10 15,clip]{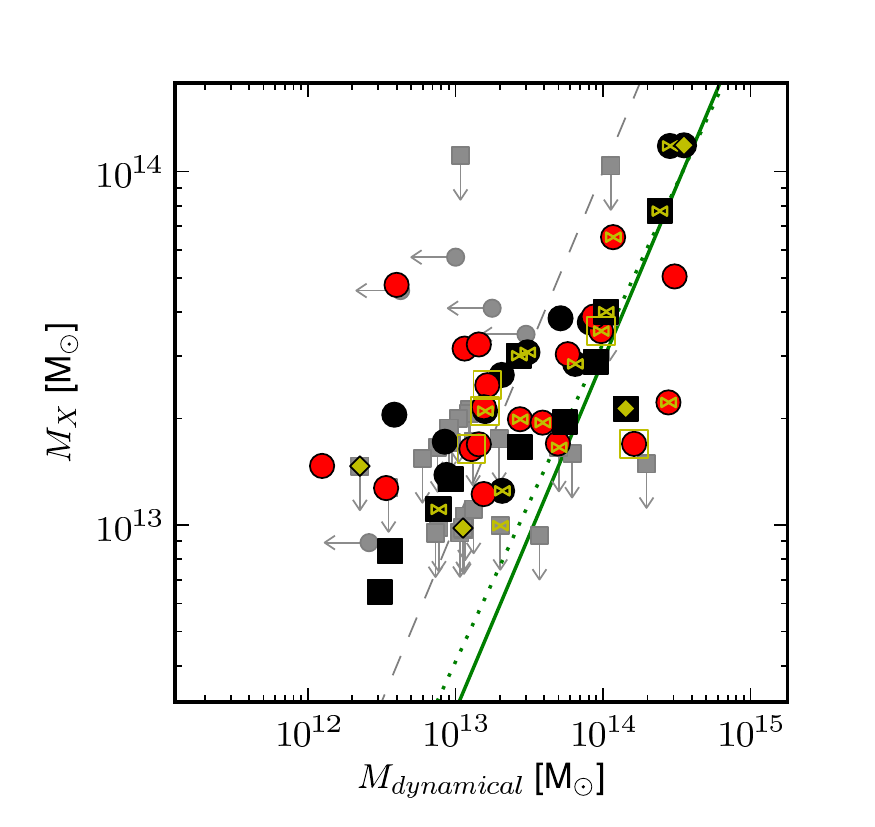}
\caption{M$_{X}$-M$_{dyn}$ relation for quality 1 \& 2 X-ray (black and red circles respectively) and optical systems (black squares) within r$_{\textrm{\tiny200,X}}$.  Grey arrows indicate limits.  Yellow bow-ties show systems tested for substructure.  Filled yellow diamonds and squares indicate systems with substructure according to AD and DS tests respectively.  Open yellow squares show groups in X-ray confused regions.  A 1:1 line is shown in dashed grey. \citet{Girardi1998} weighted and bisecting regression lines are shown for comparison as green solid and dotted lines respectively.}
\label{fig:fig9_mxmdynx}
\end{minipage}
\end{figure}

Fig.\,\ref{fig:fig9_mxmdynx} presents the two `total' mass measures for our samples: the X-ray and dynamical mass measures.  In this figure we show an X-ray based r$_{\textrm{\tiny200}}$ radial cut but, regardless of the radial cut applied, the disagreement between these measures increases for the average group, and the scatter decreases, with increasing dynamical mass.  This is not unexpected, since the range in dynamical mass is much larger than in X-ray mass (recall Fig.\,\ref{fig:fig7_sigsig}). For $\sigma$ and 1\,Mpc radial cuts, the dynamical mass may be inflated by overestimates of velocity dispersion in systems with dynamical complexity.  In general, X-ray masses are preferable, better discerning the virialized core of the system, but, for systems undetected in X-rays, this tracer of halo mass is unavailable.  \citet{Girardi1998} find, for an inhomogeneous sample of clusters, good agreement between virial and X-ray masses. We show in Fig.\,\ref{fig:fig9_mxmdynx} their weighted regression lines for comparison.  We find our masses are less and less in agreement with increasing total mass. Dynamical masses for massive systems might be improved with better dynamical modelling (e.g. `caustic masses', e.g. \citealt{Andreon2010}, \citealt{Serra2011}, etc.), but such estimates are only possible when the number of spectroscopic galaxies in and around the group (cluster) is high.
 
%PLOT - MX-MDYN
 
 %SECTION: MASS IN STARS
 \section{Mass in Stars}
 \label{sec:minstars}

\subsection{Stellar Versus `Total' Group Mass}

\begin{figure}[!htbp]
\begin{minipage}[t]{1.\linewidth}
\includegraphics[scale=1, trim= 10 7 10 15,clip]{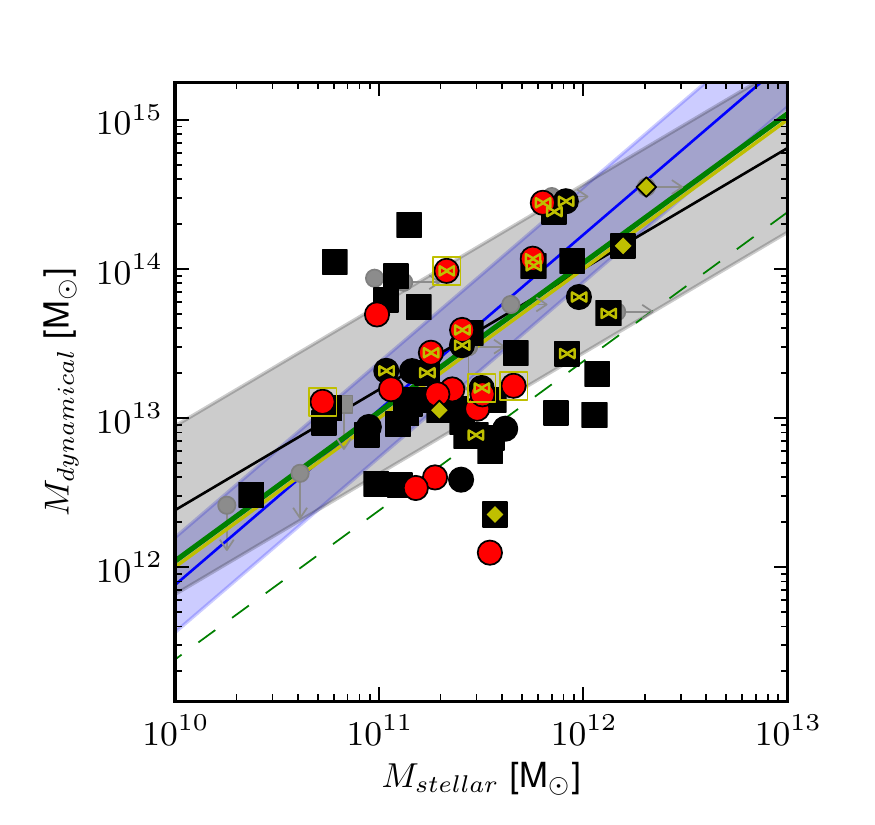}
\caption{M$_{dyn}$-M$_{stellar}$ relation for quality 1 \& 2 X-ray (black and red circles respectively) and optical systems (black squares) within r$_{\textrm{\tiny200,X}}$.  Grey arrows indicate limits.  Yellow bow-ties show systems tested for substructure.  Filled yellow diamonds and squares indicate systems with substructure according to AD and DS tests respectively.  Open yellow squares show groups in X-ray confused regions.  Bayesian best fits for the X-ray systems (blue) and optical systems (black) are shown with filled region representing the scatter.  Constant stellar mass fractions of 0.009 and 0.042 are shown in green solid and dashed lines respectively and correspond to the fractions found for M$_{\textrm{\tiny200}}$ halo masses of 14.9 and 13.7 M$_{\sun}$ by \cite{Andreon2010}.  The average 1\% fraction found by \cite{Balogh2011b} within r$_{\textrm{\tiny500}}$ for nearby low-mass clusters is shown in yellow.}
\label{fig:fig10_mdynmstell}
\end{minipage}
\end{figure}

The dynamical and group stellar masses are compared in Fig.\,\ref{fig:fig10_mdynmstell}.  Stellar mass fractions for two different group halo masses from \citet{Andreon2010} are overplotted for comparison.  Our best fits are similar to the stellar mass fraction of 0.009 found by Andreon for a 10$^{14.9}$ M$_{\sun}$ halo.  To derive total stellar masses, Andreon intregrates the total luminosity function for all red galaxies in a cluster, assuming that, in the cluster regime, blue galaxies contribute little to the overall luminosity, and assumes a dynamical M/L from \cite{Cappellari2006}.  Our results are in good agreement average 1\% stellar to dynamical mass fraction within r$_{\textrm{\tiny500}}$ found by \cite{Balogh2011b} for a sample of low-mass nearby clusters.

\begin{figure}[!htbp]
\begin{minipage}[t]{1.\linewidth}
\includegraphics[scale=1, trim= 10 7 10 15,clip]{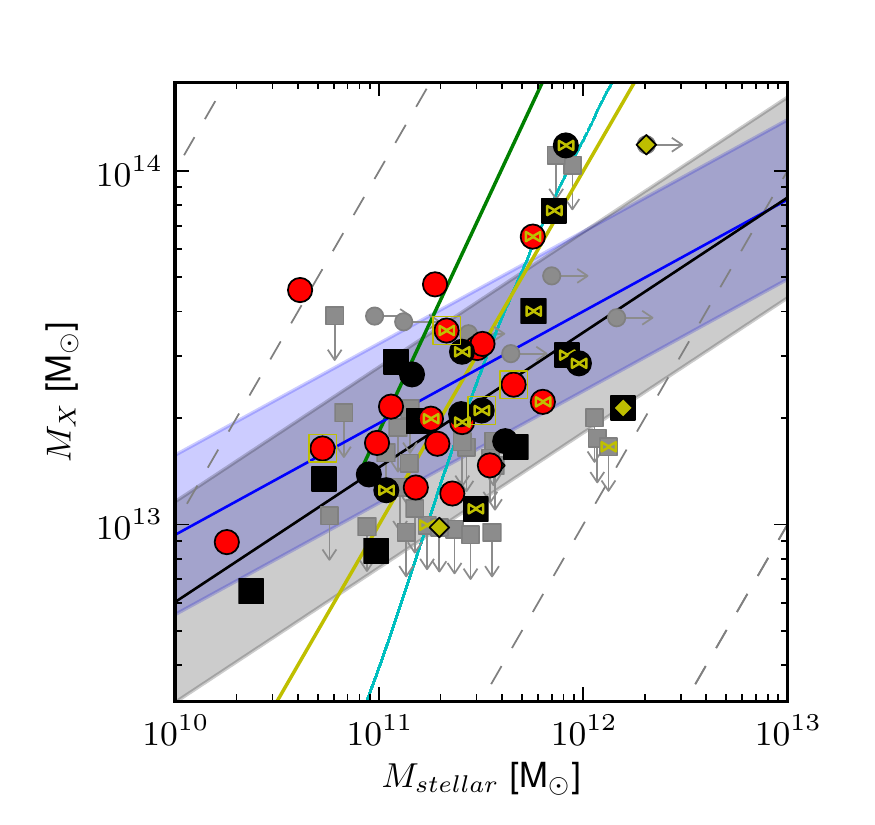}
\caption{M$_{X}$-M$_{stellar}$ relation for quality 1 \& 2 X-ray (black and red circles respectively) systems within r$_{\textrm{\tiny200,X}}$.  Grey arrows indicate limits.  Yellow bow-ties show systems tested for substructure.  Filled yellow diamonds and squares indicate systems with substructure according to AD and DS tests respectively.  Open yellow squares show groups in X-ray confused regions.  Bayesian best fits for the X-ray systems (blue) and optical systems (black) are shown with filled region representing the scatter.  \cite{Yang2009} data are over-plotted in cyan, \cite{Giodini2009} relation is shown in green, and the average 1\% fraction found within r$_{\textrm{\tiny500}}$ for nearby low-mass clusters by \cite{Balogh2011b} shown in yellow.  Lines of constant mass are shown as grey dashed lines.}
\label{fig:fig11_mxmstell}
\end{minipage}
\end{figure}

Fig.\,\ref{fig:fig11_mxmstell} shows the stellar and X-ray masses of all of our systems.  We compare relations found by \cite{Yang2009} for a low redshift sample of groups selected from SDSS, \cite{Giodini2009} for  0.1 $\le$ z $\le$ 1 COSMOS X-ray detected groups, and \cite{Balogh2011b} for nearby clusters and find relatively good agreement.  For the latter comparison, we shift the Giodini relation, which was computed for an r$_{\textrm{\tiny500}}$ radial cut assuming a simple conversion of M$_{\textrm{\tiny200}} \sim$ M$_{\textrm{\tiny500}}/0.7$ and shifted by 0.25 dex to account for the difference in assumed IMFs \citep[see][]{Leauthaud2011}.  Note that Giodini et al. integrate down to a stellar mass limit of 10$^{8}$ M$_{\sun}$ which, when compared to our limit of 10$^{10}$ M$_{\sun}$, means their total group stellar masses should be slightly higher.  At lower X-ray mass, our derived best linear fit indicates significantly lower stellar masses than Yang or Giodini. However, the latter notes that, in this low X-ray mass region, their stellar masses can range by a factor of 10 at a fixed total mass and that the logarithmic intrinsic scatter of their relation is of order 35\%.  Our results well match the average 1\% stellar to X-ray mass fraction found by \cite{Balogh2011b} for a sample of low-mass nearby clusters within r$_{\textrm{\tiny500}}$.

%PLOT - MDYN-MSTELL
%PLOT - MX-MSTELL

Both the M$_{dyn}$-M$_{stellar}$ and M$_{X}$-M$_{stellar}$ relations show a wide range in stellar mass for a given `total' mass especially considering the limits.  With a $\sigma$ derived r$_{\textrm{\tiny200}}$, the M$_{dyn}$-M$_{stellar}$ relation ${appears}$ to improve.  However, this is merely due to the increased range in velocity dispersion and thus M$_{dyn}$ with most high $\sigma$ groups exhibiting dynamical complexity (recall Fig.\,\ref{fig:fig8_lxsig}).  This drives up M$_{dyn}$ but can also increase M$_{stellar}$ due to the increased membership resulting from larger r$_{\textrm{\tiny200}}$. Therefore we choose not to show the $\sigma$ derived r$_{\textrm{\tiny200}}$ cut version of this relation as its relative tightness is misleading. The M$_{X}$-M$_{stellar}$ relations behave similarly when comparing the differently defined radial cuts. 

\begin{figure}[!htbp]
\begin{minipage}[t]{1\linewidth}
      \includegraphics[scale=1,trim=0 10 0 10,clip]{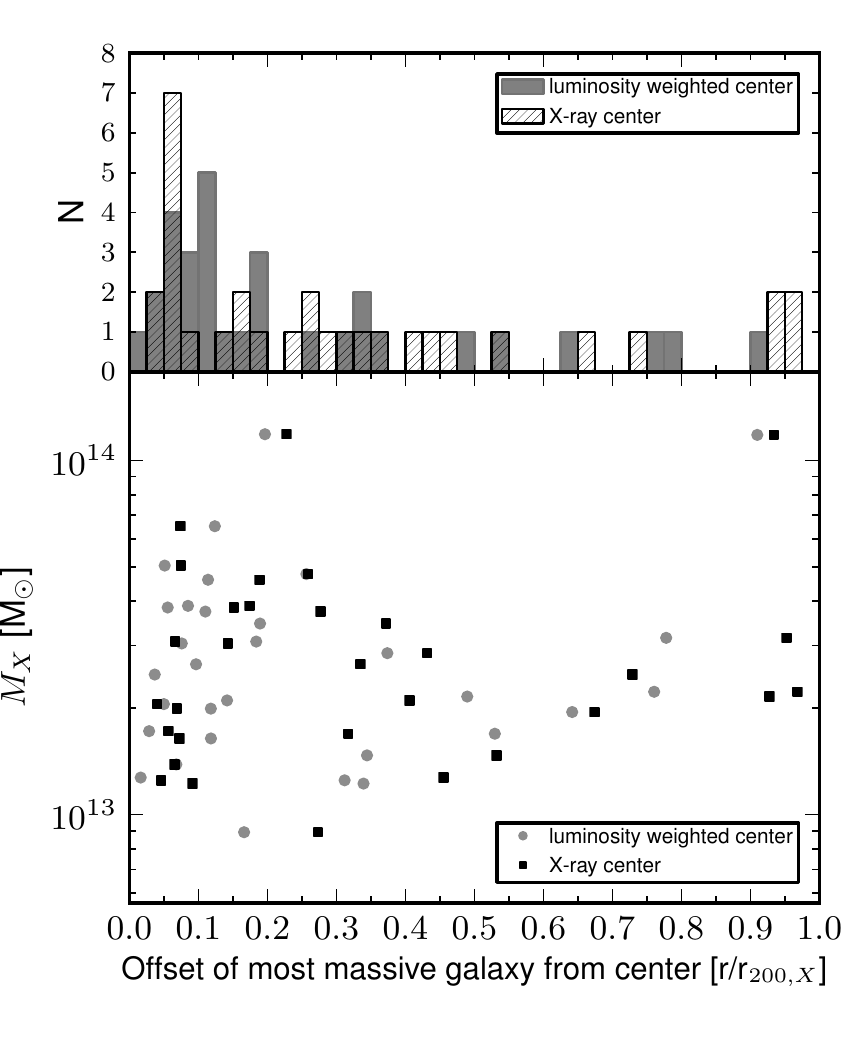}
       \caption{Histogram of the offset of the most massive galaxy from the group center (top) and versus total (X-ray) group mass (bottom). Offsets from X-ray centers are shown as black hashed histogram and black squares and luminosity weighted centers are shown as grey histogram and circles.}
     \label{fig:fig12_mmcen}
\end{minipage} 
\end{figure}

To determine if the scatter in M$_{stellar}$ given fixed total mass may be related to the dominance of the most massive galaxy (MMG), we first identify the MMG in each group and examine the offset of this galaxy from the group center. Fig.\,\ref{fig:fig12_mmcen} shows the histogram of offsets and the offset versus the total (X-ray) system mass.  The MMG generally lies near the group center regardless of whether an X-ray or luminosity weighted center is used.  With the exception of a single system, groups with higher X-ray mass (M$_{X}\gtrsim3.5\times10^{13}$ M$_{\sun}$) have the MMG within the inner third of the X-ray derived r$_{200}$. This corresponds to a distance of less than 200 kpc from the group center (nearer allowing for centering accuracy, see Fig.\,\ref{fig:fig4_cencomp}).  For lower mass systems, there is a much greater scatter in the offset of the MMG. The group with its MMG at greatest offset from the center is XR21h14 (OP21h138), a massive group with two bright stars near the (projected) center.  It is likely that the complication introduced to the photometry in this area due to the presence of these stars may be obscuring the actual MMG for this group.  The stellar mass contributed by the most massive galaxies does decrease with increasing total stellar mass, regardless of radial cut or choice of group center from an average fraction of $\sim$\,0.5 at $3 \times 10^{10}$ M$_{\sun}$ to $\lesssim$\,0.2 at $2 \times 10^{12}$ M$_{\sun}$.  In the cluster regime, \cite{Sanderson2009} found that the offset of the BCG relates both to activity in that galaxy and to the dynamical state of the cluster itself.  A full study of our BGGs (or MMGs) which includes correlation of offsets with emission properties as in Sanderson and also exploration of issues such as multiple component BGGs \citep[as in][]{Jeltema2007} would be an interesting addition to future work on galaxy properties.

 %PLOT - MOST MASSIVE GALAXY AND GROUP CENTER COMPARISON

\subsection{Stellar Mass Fractions}

Using the X-ray mass to represent the total halo mass, our best fits for both the high quality X-ray and optically selected systems indicate almost constant fractions (M$_{stellar}$/M$_{X}$) of $\sim$0.011, independent of halo mass (M$_{X}$).  This is different from the $mean$ fraction M$_{stellar}$/M$_{X}$$\sim$0.014 with logarithmic standard deviation of 0.398. Recall that we integrate our stellar mass down to $10^{10}$  M$_{\sun}$.  Using dynamical mass instead as the total group mass results in a mean fraction of $\sim$0.022. Examining the M$_{dyn}$-M$_{stellar}$ best fits, the fraction is similar to the $\sim$0.009  found by Andreon for a 10$^{14.9}$ M$_{\sun}$ halo. Giodini et al. find that the stellar mass fraction associated with galaxies within r$_{\textrm{\tiny500}}$ decreases with increasing total mass as M$_{\textrm{\tiny500}}^{-0.26\pm0.04}$.  We do not find a similar relation but note that the relation found by Giodini et al. breaks down when clusters are excluded.

\begin{figure}[!tbp]
\begin{minipage}[t]{1\linewidth}
      \vspace{0pt}
      \includegraphics[scale=1, trim=0 10 0 20,clip]{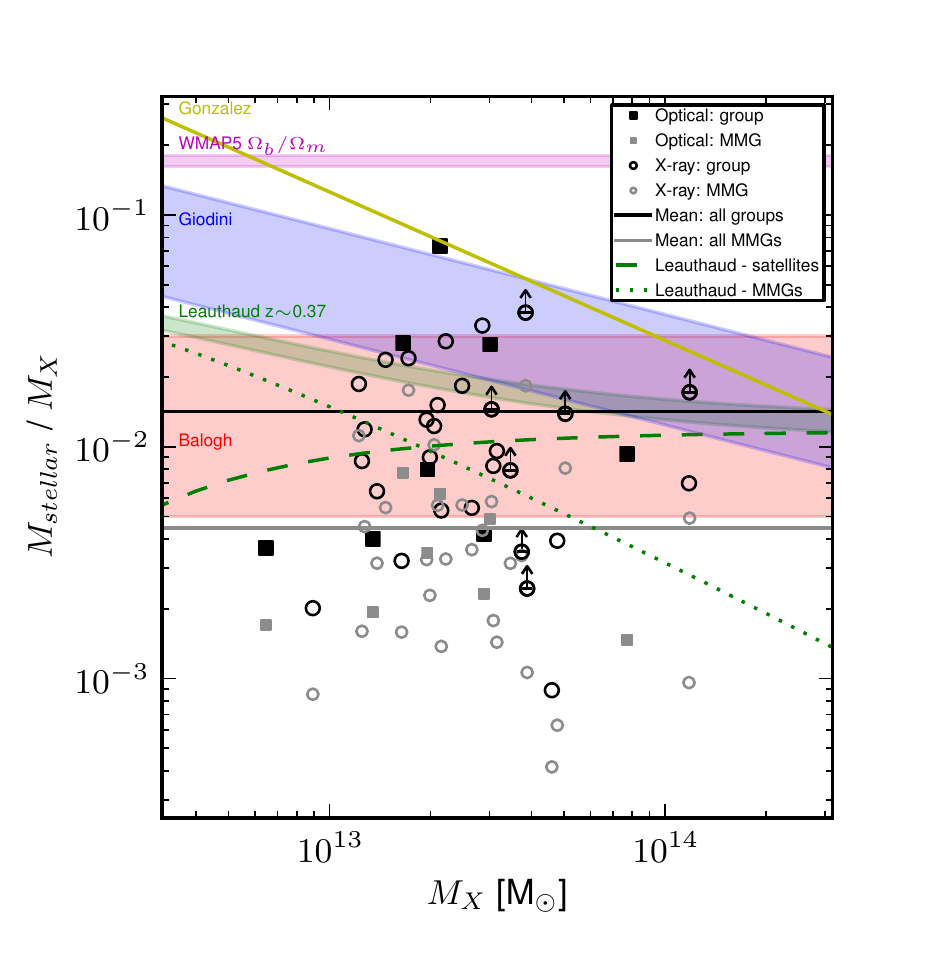}
       \caption{Stellar mass fraction versus total mass.  Black and grey circles indicate total and most massive galaxy fractions respectively for X-ray systems while squares indicate similar quantities for optical systems.   The mean stellar mass fraction for our entire sample (high quality X-ray and optical samples) is indicated by a black solid line and that of the MMG in grey.  The \cite{Leauthaud2011} z$\sim$0.37 stellar mass fraction is shown in green with dotted green line indicating central and dashed green line indicating the contribution from satellite galaxies. The \cite{Giodini2009} COSMOS sample stellar mass fraction and its intrinsic scatter is shown in blue. The \cite{Balogh2011b} fraction measured within r$_{\textrm{\tiny500}}$ is shown in red and the \cite{Gonzalez2007} in yellow.  The baryon fraction from WMAP5 \citep{Dunkley2009} is plotted in magenta.}
     \label{fig:fig13_stellfrac}
\end{minipage} 
\end{figure}

Fig.\,\ref{fig:fig13_stellfrac} shows the stellar mass fraction versus the total (X-ray) system mass given an X-ray based r$_{\textrm{\tiny200}}$ cut for the total stellar mass and mass of the most massive galaxy. The mean contribution of the most massive galaxy to the total system mass (M$^{\textrm{\tiny{MMG}}}_{stellar}$/M$_{X}$) including both quality 1 \& 2 X-ray and optically selected systems is $\sim$0.004.  In addition to constant lines approximating our mean stellar mass fractions for an X-ray based halo mass, we show the fractions found by \cite{Giodini2009} for their COSMOS sample of groups, \citet{Balogh2011b} within r$_{\textrm{\tiny500}}$ for their nearby low-mass galaxy clusters, the \cite{Leauthaud2011} z$\sim$0.37 COSMOS derived sample of groups and clusters, and the fraction including intracluster light (ICL) within r$_{\textrm{\tiny500}}$ from the \cite{Gonzalez2007} cluster sample.  We again shift the Giodini relation to account for a difference in IMF.  Additionally, the fractions found by Leauthaud by dividing the group population into central and satellite galaxies are shown.  In the cluster regime, stellar mass fractions are not universal, but  generally decrease with increasing cluster mass \citep[e.g.][]{Ramella2004,Eke2005,Giodini2009}. We find our fraction to be significantly lower than the fraction including ICL found by \cite{Gonzalez2007} with this divergence increasing with decreasing X-ray mass. We find many more systems at lower stellar mass fractions than do Giodini et al.  Our spectroscopic selection results in a larger scatter in the stellar mass fraction. 

%PLOT - STELLAR MASS FRACTIONS AND MOST MASSIVE GALAXY

%SECTION: UNDERLUMINOUS GROUPS
\section{Underluminous Groups}
\label{sec:underlum}

\begin{SCfigure*}[0.7][t]
\centering
\includegraphics[scale=0.88, trim = 2 10 25 15,clip]{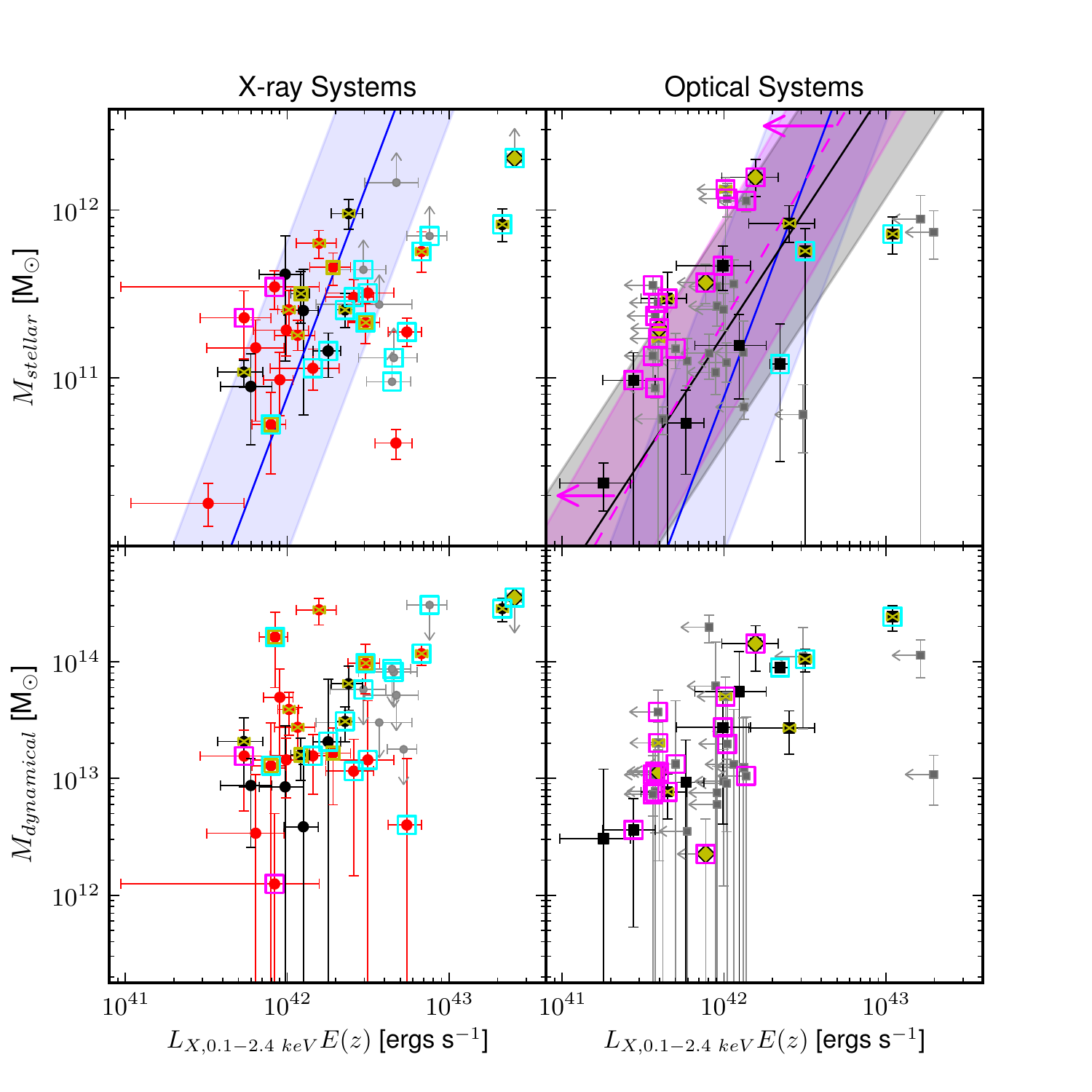}
\caption{\\ \\ TOP: M$_{stellar}$-L$_{X}$ relation for X-ray selected systems (left) and optical systems (right) with X-ray based r$_{\textrm{\tiny200}}$ cut applied. Bayesian best fits are shown in blue and black (X-ray and optical fits respectively) with filled regions representing the scatter. Bayesian best fit for optical systems where upper limits on X-ray luminosity have been treated as detections is shown in magenta with filled region representing the scatter and magenta arrows reflecting that this relation is in reality likely shifted to lower X-ray luminosities. \\ \\ \\ BOTTOM: M$_{dyn}$-L$_{X}$relation for X-ray (left) and optically (right) selected systems with X-ray based r$_{\textrm{\tiny200}}$ cut applied. Quality 1 \& 2 X-ray selected systems are shown as black and red circles respectively while optical systems are shown as black squares.  Grey arrows indicate limits. Yellow bow-ties show systems tested for substructure.  Filled yellow diamonds and squares indicate systems with substructure according to AD and DS Tests respectively.  Open yellow squares show groups in X-ray confused regions.  Open magenta and cyan squares indicate X-ray underluminous and overluminous systems respectively.\\ \\ \\ \\ \\ \\ }
\label{fig:fig14_lxmstellmdyn}
\end{SCfigure*}

Fig.\,\ref{fig:fig14_lxmstellmdyn} (top panel) shows the M$_{stellar}$-L$_{X}$ relations for X-ray and optical groups respectively with an X-ray based r$_{\textrm{\tiny200}}$ cut applied.  We provide the best fit  slope, intercept, and intrinsic scatter, and uncertainties, for the reciprocal relation (L$_{X}$-M$_{stellar}$) for the optical and high quality X-ray systems for each of the different radial cuts in Tab.\,\ref{tab:lxmstellfits}.  Note that errors in stellar mass are averaged in order to produce symmetric errors for input into the Kelly Bayesian best fit procedure. The best fit solutions for X-ray systems vary significantly between the quality 1 and quality 1 \& 2 systems and between different radial cuts.  Considering only the quality 1 X-ray systems, the 1\,Mpc radial cut produces a relatively shallow relation, similar to that for optical and Q = 1 \& 2 X-ray system considering the very large intrinsic scatter, but both r$_{200}$ based cuts for these highest quality systems are significantly steeper.  Comparing the quality 1 \& 2 X-ray systems only to that for the optical groups and taking the uncertainties into account, the best fit to the relations for both samples are similar.  If the upper limits in X-ray luminosity were included, assuming the luminosity is the value of the limit (the maximum possible), the optical groups would be on average comparatively {\it underluminous} in X-rays.

Including these limits, we perform the best fit again for our systems, splitting the entire population, including all optical systems and Q = 1 \& 2 X-ray systems, into three types of groups: `underluminous', `normal', and `overluminous' {\it relative to their stellar mass}.  We define X-ray under- and overluminous groups as those which have lower/higher L$_{X}$ than the best fit value (including upper limits), minus/plus half the scatter (0.5$s$).  Those groups that are underluminous, having higher stellar masses and lower luminosities than the fit even including half the scatter, are marked by open magenta squares while the overluminous systems are marked similarly in cyan.  Note that while groups with upper limits in X-ray luminosity are included in the underluminous group population, we exclude these from the overluminous subset as they may in fact be consistent with the relation.

Note that to further ensure the robustness of these results, the under- and overluminous groups were also defined relative to the best fit based on the {\it X-ray} (Q = 1 \& 2) systems and, though specific numbers changed, all qualitative results remained consistent.  However, using only the Q=1 groups for either of the r$_{200}$ based cuts would result in a more cluster-like slope of the L$_{X}$-M$_{stellar}$ relation (top left panel of Fig.\,\ref{fig:fig14_lxmstellmdyn}) and a significant difference in the population of groups defined as under- or overluminous.  

Several of the outliers to the M$_{stellar}$-L$_{X}$ relation, including an underluminous optical and the most overluminous X-ray group -- which are also the systems with the highest stellar mass -- show substructure.  However, some of the groups with $\geq$\,10 members (marked as yellow bow-ties) do not show evidence of substructure from either the AD or DS test and are among the most significant outliers from this relation.

%MSTELL-LX BEST FITS
\tabletypesize{\scriptsize}
\setlength{\tabcolsep}{0.0275in}
\begin{table}[!tbp]
\begin{center}
\caption{ L$_{\textrm{x}}$-M$_{\textrm{stellar}}$ Relation Bayesian Best Fits}
\begin{tabular*}{0.5\textwidth}{l l|c|c|c|}
\cline{3-5}
& & $m$ & $c$ & $s$ \tstrut \\ [2pt]
\cline{1-5}
\multirow{3}{*}{X-ray Q=1}
& 1Mpc & 0.2506$\pm^{0.9341}_{0.8836}$ & 39.184$\pm^{10.542}_{11.047}$ & 0.6271$\pm^{0.3061}_{0.1676}$  \tstrut \\ [1.5pt] \cline{2-5}
& r$_{\textrm{\tiny{200,}\textrm{\scriptsize{$\sigma$}}}}$ & 1.0128$\pm^{0.4214}_{0.3992}$ & 30.463$\pm^{4.6178}_{4.9364}$ & 0.3981$\pm^{0.2403}_{0.1378}$  \tstrut \\ [1.5pt] \cline{2-5}
& r$_{\textrm{\tiny{200,X}}}$ & 1.0659$\pm^{0.7493}_{0.7725}$ & 29.847$\pm^{8.8896}_{8.7808}$ & 0.5017$\pm^{0.2429}_{0.1500}$  \tstrut \\ [1.5pt]
\cline{1-5}
\multirow{3}{*}{X-ray Q=1 \& 2}
& 1Mpc & 0.1636$\pm^{0.1822}_{0.1759}$ & 40.301$\pm^{2.0875}_{2.1560}$ & 0.4057$\pm^{0.0761}_{0.0582}$  \tstrut \\ [1.5pt] \cline{2-5}
& r$_{\textrm{\tiny{200,}\textrm{\scriptsize{$\sigma$}}}}$ & 0.4285$\pm^{0.1563}_{0.1564}$ & 37.197$\pm^{1.8092}_{1.8303}$ & 0.3228$\pm^{0.0694}_{0.0552}$  \tstrut \\ [1.5pt] \cline{2-5}
& r$_{\textrm{\tiny{200,X}}}$ & 0.3903$\pm^{0.1779}_{0.1748}$ & 37.749$\pm^{1.9910}_{2.0485}$ & 0.3625$\pm^{0.0647}_{0.0544}$  \tstrut \\ [1.5pt]
\cline{1-5}
\multirow{2}{*}{Optical}
& 1Mpc & 0.5915$\pm^{0.3093}_{0.2970}$ & 35.288$\pm^{3.4474}_{3.5627}$ & 0.4661$\pm^{0.1646}_{0.1085}$  \tstrut \\ [1.5pt] \cline{2-5}
& r$_{\textrm{\tiny{200,}\textrm{\scriptsize{$\sigma$}}}}$ & 0.5266$\pm^{0.2603}_{0.2653}$ & 36.093$\pm^{3.0918}_{3.0718}$ & 0.4010$\pm^{0.1570}_{0.1014}$  \tstrut \\ [1.5pt] \cline{2-5}
& r$_{\textrm{\tiny{200,X}}}$ & 0.7034$\pm^{0.2943}_{0.3040}$ & 34.042$\pm^{3.4661}_{3.3378}$ & 0.4459$\pm^{0.1849}_{0.1188}$  \tstrut \\ [1.5pt]
\cline{1-5}
\multirow{2}{*}{Optical with}
& 1Mpc & 0.5161$\pm^{0.1802}_{0.1841}$ & 36.081$\pm^{2.1159}_{2.0891}$ & 0.3870$\pm^{0.0648}_{0.0521}$  \tstrut \\ [1.5pt] \cline{2-5}
\multirow{2}{*}{upper limits}
& r$_{\textrm{\tiny{200,}\textrm{\scriptsize{$\sigma$}}}}$ & 0.5198$\pm^{0.1678}_{0.1658}$ & 36.089$\pm^{1.9194}_{1.9111}$ & 0.3640$\pm^{0.0679}_{0.0566}$  \tstrut \\ [1.5pt] \cline{2-5}
& r$_{\textrm{\tiny{200,X}}}$ & 0.5956$\pm^{0.1754}_{0.1789}$ & 35.243$\pm^{2.0368}_{1.9962}$ & 0.3691$\pm^{0.0687}_{0.0545}$  \tstrut \\ [1.5pt]
\cline{1-5}
\end{tabular*}
\label{tab:lxmstellfits}
\end{center}
\tnote[]{Column description: Bayesian best fit slope ($m$) and lower and upper errors (column 1); intercept ($c$) and lower and upper errors  (2); and intrinsic scatter ($s$) and lower and upper errors (3) of the relation $\log$(L$_{X}$) $= m \times \log($M$_{stellar}) + c + \epsilon$, where $\epsilon$ is a random variable with variance equal to $s^2$.}
\end{table}

%PLOT

The bottom panel of Fig.\,\ref{fig:fig14_lxmstellmdyn} shows the M$_{dyn}$-L$_{X}$ relations for X-ray and optical groups respectively with an X-ray based r$_{\textrm{\tiny200}}$ cut applied.  Examining the positions of the underluminous systems, indicated by open magenta squares, their dynamical masses are not unusually low, spanning a wide range in M$_{dyn}$, and many of these underluminous groups do not exhibit significant dynamical complexity.  Recall that the latter may lead to elevated velocity dispersions and overestimation of dynamical and group stellar mass.   Groups with low X-ray luminosity relative to their stellar mass then do not exhibit particularly unusual dynamical characteristics.  This may suggest a population of dynamically young groups which are just in the process of collapse.  

Next we examine the median contribution of the MMG to the total group stellar mass (M$^{\textrm{\tiny{MMG}}}_{stellar}$/M$_{stellar}$).  Recall that M$_{stellar}$ for all groups has had substantial incompleteness corrections applied.  In order to best determine the statistical contribution from the MMG, given that we may have missed some of these galaxies, we re-calculate the group stellar mass {\it excluding} the MMG, and use the difference between this value and that found for the group including all members to characterize the fraction M$^{\textrm{\tiny{MMG}}}_{stellar}$/M$_{stellar}$.  For underluminous systems, the median contribution of the MMG to the total group stellar mass (M$^{\textrm{\tiny{MMG}}}_{stellar}$/M$_{stellar}$) is lower ($\sim\,$36$\%$) than that found for all systems ($\sim$\,42$\%$) with the most underluminous systems having less of their mass contributed from this member. 

To test the significance of the difference in the contribution of the MMG to the total stellar mass between underluminous and the total population of groups, we first create a sample matched in group stellar mass to our underluminous groups from the complete sample of X-ray and optical systems.  This process is repeated 10,000 times, calculating M$^{\textrm{\tiny{MMG}}}_{stellar}$/M$_{stellar}$ for each group in each sample. Finally, we calculate for each sample the number of systems having $\mathrm{M}^{\textrm{\tiny{MMG}}}_{stellar}/\mathrm{M}_{stellar}< 40$\% (above the peak of the underluminous distribution).  For the underluminous groups, this is 80\% of 15 groups. Only 89 of the 10,000 matched samples meet this criterion -- i.e. having $\geq$\,80\% of groups with $\mathrm{M}^{\textrm{\tiny{MMG}}}_{stellar}/\mathrm{M}_{stellar}< 40$\% -- indicating that the difference is indeed significant.   This may imply that in the underluminous systems less IGM is available from relatively equal mass progenitors (which leads to the group not having a single dominant galaxy).  The existence and possible origins of X-ray underluminous or `dark' systems remains a topic of vigorous debate even in the cluster regime where X-ray and spectroscopic data are abundant.  A recent study of the maxBCG clusters and an X-ray bright subsample of the clusters by the Planck Collaboration (\citeyear{Planck2011}) finds evidence for a possible X-ray underluminous population which shows a low Sunyaev-Zel'dovich signal normalization, while \citet{AndreonMoretti2011} find no evidence for a significant population of underluminous systems in a study of X-ray luminosity in color selected clusters.  The \cite{Rykoff2008} comparison of X-ray and optically selected clusters suggests there is a wide range of L$_{X}$ at fixed mass, and that X-ray selection simply picks off the more X-ray luminous part of the population.

The overall fraction of gas scales with halo mass, with clusters having a higher gas mass fraction than groups \citep[e.g.][]{Sun2009,Pratt2009,Giodini2009,Peeples2011}.  To explore why groups with similar total stellar mass may have lower gas mass and a lower contribution of stellar mass from the most massive member, we contrast two modes of group assembly.  In the first scenario, the group begins with a massive galaxy and accretes mass smoothly.  In the second, roughly equivalent mass `subgroups' (clumps) comprised of similar mass / luminosity galaxies merge.  The former case would result in a group with both a higher gas fraction and a more massive central galaxy.  We posit this may be one explanation for the observed correlation between the fraction of mass in the most massive galaxy, and the relative X-ray luminosity.  \cite{Popesso2007} use an optically selected cluster sample to explore the nature of underluminous systems, finding evidence that these systems are undergoing a phase of mass accretion and are still accreting intracluster gas or in the process of merging.  In the future, it would be interesting to examine the galaxy population in our groups to examine the role of the X-ray emitting hot medium in driving galaxy evolution in this mass regime.

Examining the positions of the optically and X-ray selected overluminous systems on the M$_{dyn}$-L$_{X}$ relation (bottom panel of Fig.\,\ref{fig:fig14_lxmstellmdyn}), indicated by open cyan squares, the dynamical masses of the former are all very high ($\gtrsim$\,10$^{14}$ M$_{\sun}$) while the latter span the entire range in M$_{dyn}$. The median contribution of the MMG to the total group stellar mass (M$^{\textrm{\tiny{MMG}}}_{stellar}$/M$_{stellar}$) is higher ($\sim$\,47$\%$) for overluminous systems than for all systems ($\sim$\,42$\%$).  As for the underluminous groups, we test the significance of this difference in the relative contribution from the MMG by creating a sample matched in mass to our overluminous groups from the complete sample of X-ray and optical systems.  In this case, we calculate for each sample the number of systems having M$^{\textrm{\tiny{MMG}}}_{stellar}$/M$_{stellar}$$>$ 50\%. For the overluminous groups, this is 47\% of 17 groups. 845 of the 10,000 matched samples meet this criterion, having $\geq$\,47\% of groups with M$^{\textrm{\tiny{MMG}}}_{stellar}$/M$_{stellar}$$>$ 50\%, indicating that this difference is not as significant as that found for underluminous systems.

%SECTION: CONCLUSIONS
\section{Conclusions}
\label{sec:conclusions}

We have defined two group samples at 0.12$<$z$<$0.79 in the same fields, one containing 39 high quality X-ray selected systems and the other 38 optically selected systems, in order to study groups spanning a significant mass and evolutionary range. Group membership was defined and we applied three different radial cuts: two r$_{\textrm{\tiny200}}$ cuts (roughly approximating a virial radius) based on the X-ray emission and velocity dispersion of the systems; and a constant 1\,Mpc cut.  Group masses were estimated from X-ray and dynamical characteristics and stellar content -- the latter two within the differing radial cuts.  Dynamical complexity and substructure was explored using the Anderson-Darling and Dressler-Schectman tests and the shape of X-ray emission. We presented the L$_{X}$-$\sigma$ relation for our systems which is similar to that found for nearby groups and discussed the effects of centering, radial cuts and dynamical complexity/substructure in regards to outliers in this, and other scaling relations.  Best fits to this, and to L$_{X}$-M$_{stellar}$ relations for different group samples and radial cuts were presented.  Stellar mass fractions were estimated using the X-ray and dynamical mass as proxies for the group halo mass.  Finally, evidence for a population of optical systems seemingly underluminous in X-rays given their stellar and dynamical mass was discussed. Our main conclusions are as follows:\\

\noindent $\bullet$ Confusion:\\[3pt]
Confusion exists both in matching galaxies to extended X-ray emission and matching X-ray emission to already identified optical systems.  Until X-ray spectroscopy is available to measure the redshift of the X-ray emitting gas, completely confident matching will not be possible.  Splitting systems into X-ray detected and undetected systems designates the problem, not the solution.  These difficulties in matching make cosmological studies using groups difficult.  \\

\noindent $\bullet$ Dynamical complexity:\\[3pt]
Dynamical complexity/substructure in a system \textit{can} work to inflate velocity dispersion and stellar mass and may explain the position of certain outliers in the scaling relations explored here.  It is important to recall that the tests we are using are orbit dependent and can only be confidently applied to systems having at least ten members.\\

\noindent $\bullet$ Radial cuts:\\[3pt]
Applying X-ray based r$_{\textrm{\tiny200}}$ radial cuts usually produces the tightest scaling relations.  The good correlation between L$_{X}$ and $\sigma$ and the lack of dynamical complexity found for systems using this radius implies that it is isolating the virialized part of the group.  Velocity dispersion based and constant cuts generally result in larger radii, more members, and include more substructure/non-Gaussianity.  This acts to increase scatter and inflate both velocity dispersion and stellar mass. However, as some systems are not X-ray detected, such cuts are the only options.\\

\noindent $\bullet$ Stellar mass fraction:\\[3pt]
We find a mean stellar mass fraction of $\sim$0.014 within an X-ray based r$_{\textrm{\tiny200}}$ and treating the X-ray mass as the total mass of the system.  This is comparable to those found by \cite{Giodini2009}, \cite{Balogh2011b} and \cite{Leauthaud2011} but significantly lower than that found by \cite{Gonzalez2007}.  The mean contribution of the most massive galaxy is $\sim$0.004.  Using a total mass based on dynamical mass would result in different fractions due to significant disagreement between M$_{X}$ and M$_{dyn}$ for many of our systems. \\

\noindent $\bullet$ Total mass measures:\\[3pt]
The differences in total mass measures (M$_{X}$ and M$_{dyn}$) tend to increase, and the scatter decrease, as X-ray mass increases. \\

\noindent $\bullet$ X-ray underluminous groups:\\[3pt]
We define a sample of systems as X-ray underluminous given their stellar mass, the majority of which are optically selected.  Not all such systems show dynamical complexity and the stellar mass fraction in the most massive galaxy of these systems is on average less than that found for the total population of groups.   This may indicate that less IGM is being contributed from the progenitor halo containing the most massive member and we posit that differences in accretion (a continuous smooth accretion of galaxies from the field verses the merging of similar mass `subgroups')  may be one explanation for the observed correlation between the fraction of mass in the most massive galaxy and the relative X-ray luminosity.
\\

\acknowledgments
We would like to thank the CNOC2 team for the use of their unpublished data.  This research uses observations made with ESO Telescopes at Paranal (program ID 080.A-0427 and 081.A-0103) and the Magellan telescopes operated by The Carnegie Institution of Washington. We acknowledge the use of the FORS and COSMOS pipelines and thank Dr. Carlo Izzo at ESO for pipeline assistance and Renbin Yan for providing the ZSPEC software.  We thank Russell Blackport for the reduction of and measurements of redshifts from GMOS data.  LCP acknowledges support from an NSERC Discovery grant. JLC thanks Francesco Montesano, Stefania Giodini, Marisa Girardi, and the referee for valuable input.\\


%% in parentheses, after the keyword, but they will not be verified.
%% can look these up here! http://dopey.mcmaster.ca/facilities/facilities.html

{\it Facilities:} \facility{VLT: (FORS2)}, \facility{Magellan: Baade (LDSS2 imaging spectrograph)}, \facility{CFHT (WIRCAM)}, \facility{NTT (SOFI)}, \facility{XMM}, \facility{CXO}, \facility{GALEX}

\bibliography{abbrevs,short_abbrevs,bib_globalgrpprops_final}

%==================================================================================
%TABLES
%==================================================================================

\clearpage
\pagebreak

\setcounter{table}{1}
\setlength{\tabcolsep}{0.05in} 

\LongTables

\setlength{\tabcolsep}{0.1in} 
%{\footnotesize
\begin{longtable}{lllllll}
\caption[Spectroscopic Redshifts]{Spectroscopic Redshifts} \label{Schectman} \\
\hline \hline \\[-2ex]
RA & Dec & z & source & 1 Mpc & r$_{200,\sigma}$ & r$_{200,X}$ \\   J2000 & J2000 & & & Group Member & Group Member  & Group Member \\ \hline
\endfirsthead
RA & Dec & z & source & 1 Mpc  & r$_{200,\sigma}$ & r$_{200,X}$ \\ \hline
\endhead
221.927383 & 9.147980 & 0.5685 & GMOS & \nodata & \nodata & \nodata \\
221.941650 & 9.159590 & 0.8850 & GMOS & \nodata & \nodata & \nodata \\
221.959946 & 9.192980 & 0.8497 & GMOS & \nodata & \nodata & \nodata \\
221.963501 & 9.136121 & 0.8112 & GMOS & \nodata & \nodata & \nodata \\
221.964722 & 9.137767 & 0.8116 & GMOS & \nodata & \nodata & \nodata \\
221.981888 & 9.151252 & 0.7233 & GMOS & \nodata & \nodata & \nodata \\
221.993881 & 9.211152 & 0.7040 & GMOS & \nodata & \nodata & \nodata \\
222.003189 & 9.144666 & 0.8836 & GMOS & \nodata & \nodata & \nodata \\
222.014923 & 9.215993 & 0.7014 & GMOS & \nodata & \nodata & \nodata \\
222.019150 & 8.903519 & 0.3581 & CNOC2 & \nodata & \nodata & \nodata \\
\hline
\end{longtable}
\normalsize {This table is published in its entirety in the electronic edition. A portion is shown here for guidance regarding its form and content.  Column description: object R.A. and Decl. for Equinox J2000.0 (columns 1 \& 2); object redshift (3); source of object redshift (4); X-ray and/or optical group(s) for which object is a member for 1\,Mpc, r$_{200,\sigma}$, \& r$_{200,X}$ radial cuts (5,6,7).}
%}

\clearpage
\pagebreak

\LongTables

\tabletypesize{\tiny}
\begin{deluxetable*}{lllccccc}
 % \tablenum{2}
  \tablecolumns{8}
  \tablecaption{Basic Properties of X-ray Selected Groups \label{tab:Xgrp_props1}}
  \tablehead{\colhead{XID} & \colhead{IAU name} & \colhead{RA} & \colhead{Dec} & \colhead{z} & \colhead{Q} & \colhead{X-ray Flux} & \colhead{X-ray Significance} \\ & & \colhead{J2000} & \colhead{J2000} & & & \colhead{ 10$^{\textrm{\tiny{-14}}}$ ergs cm$^{\textrm{\tiny{-2}}}$}}
\startdata
XR14h02 & 144949+0910.9 & 222.45584 & +9.18167 & 0.1315 & 1.0 & 0.89$\pm$0.31 & 2.84 \\ [2pt]
XR14h09 & 145009+0904.3 & 222.54075 & +9.07188 & 0.6420 & 1.0 & 1.23$\pm$0.12 & 9.75 \\ [2pt]
XR14h10 & 144905+0905.5 & 222.27293 & +9.09198 & 0.7382 & 1.0 & 0.13$\pm$0.05 & 2.48 \\ [2pt]
XR21h08 & 215143$-$0526.0 & 327.93324 & $-$5.43371 & 0.2617 & 1.0 & 0.77$\pm$0.10 & 7.74 \\ [2pt]
XR21h10 & 215045$-$0530.9 & 327.68966 & $-$5.51535 & 0.3383 & 1.0 & 0.17$\pm$0.05 & 3.23 \\ [2pt]
XR21h11 & 215111$-$0535.6 & 327.79834 & $-$5.59362 & 0.4762 & 1.0 & 0.20$\pm$0.04 & 4.5 \\ [2pt]
XR21h12 & 215136$-$0535.8 & 327.90274 & $-$5.59811 & 0.2195 & 1.0 & 0.25$\pm$0.08 & 3.2 \\ [2pt]
XR21h14 & 215041$-$0541.0 & 327.67186 & $-$5.68486 & 0.4379 & 1.0 & 2.34$\pm$0.08 & 26.24 \\
\enddata 
\tablenotetext{}{Column description: group identification number for the X-ray system (column 1); IAU name (2); R.A. and Decl. of the center of the extended X-ray emission for Equinox J2000.0 (3 \& 4); spectroscopic redshift (5); group redshift quality (6); the total flux in the 0.5$-$2 keV band (7); and significance of the X-ray flux (8)}
\end{deluxetable*}  

\clearpage
\pagebreak
\LongTables

\tabletypesize{\tiny}
\setlength{\tabcolsep}{0.025in}
\begin{deluxetable*}{l|ccccc|ccccc|ccccc}
  \tablecolumns{16}
  \tablecaption{Measured Properties of X-ray Selected Groups \label{tab:Xgrp_props2}}
  \tablehead{\colhead{XID} & \colhead{N} & \colhead{Radius} & \colhead{$\sigma$} & completeness & dynamical & \colhead{N} & \colhead{Radius} & \colhead{$\sigma$} & completeness & dynamical & \colhead{N} & \colhead{Radius} & \colhead{$\sigma$} & completeness & dynamical \\ & \colhead{1Mpc} & \colhead{[$\arcsec$]} & \colhead{[km s$^{\textrm{\tiny{-1}}}$]} & R<22 & complexity & \colhead{r$_{\textrm{\tiny200,}\textrm{\scriptsize{$\sigma$}}}$} & \colhead{[$\arcsec$]} & \colhead{[km s$^{\textrm{\tiny{-1}}}$]} & R<22 & complexity & \colhead{r$_{\textrm{\tiny200,X}}$} & \colhead{[$\arcsec$]} & \colhead{[km s$^{\textrm{\tiny{-1}}}$]} & R<22 & complexity}
 \startdata
XR14h02 & 12 & 455.97 & 213$\pm$33 &  0.61 &-& 7 & 164.88 & 166$\pm$58 &  0.63 &-& 7 & 204.84 & 166$\pm$58 &  0.63 &-\\ [2pt]
XR14h09 & 15 & 155.01 & 850$\pm$112 &  0.67 &AD& 20 & 196.65 & 782$\pm$88 &  0.67 &AD,DS& 13 & 117.72 & 817$\pm$141 &  0.67 &AD\\ [2pt]
XR14h10 & 5 & 146.62 & 458$\pm$184 &  0.93 &-& 5 & 103.01 & 458$\pm$184 &  0.94 &-& 4 & 73.08 & 483$\pm$308 &  1.00 &-\\ [2pt]
XR21h08 & 14 & 263.99 & 262$\pm$38 &  0.37 &-& 9 & 135.30 & 253$\pm$54 &  0.45 &-& 11 & 147.96 & 280$\pm$46 &  0.41 &-\\ [2pt]
XR21h10 & 11 & 221.08 & 241$\pm$35 &  0.62 &-& 4 & 70.40 & 164$\pm$190 &  0.67 &-& 4 & 99.00 & 164$\pm$190 &  0.70 &-\\ [2pt]
XR21h11 & 14 & 179.72 & 374$\pm$75 &  0.61 &AD& 11 & 133.64 & 415$\pm$86 &  0.68 &-& 10 & 90.72 & 428$\pm$87 &  0.69 &-\\ [2pt]
XR21h12 & 37 & 301.07 & 255$\pm$32 &  0.58 &-& 11 & 165.52 & 265$\pm$78 &  0.61 &-& 11 & 126.72 & 265$\pm$78 &  0.62 &-\\ [2pt]
XR21h14 & 46 & 188.35 & 750$\pm$73 &  0.69 &AD& 58 & 235.72 & 683$\pm$60 &  0.69 &AD& 37 & 154.44 & 704$\pm$81 &  0.67 &-\\
\enddata
\tablenotetext{}{Column description: group identification number for the X-ray system (column 1); number of member galaxies within 1\,Mpc (2); the radius in arcseconds of 1\,Mpc (3); the velocity dispersion within a 1\,Mpc cut (4); spectroscopic completeness (to R<22) within a 1\,Mpc cut  (5), dynamical complexity within a 1\,Mpc cut  (6); the number of member galaxies within a $\sigma$ derived r$_{\textrm{\tiny200}}$ (7); the radius in arcseconds of a $\sigma$ derived r$_{\textrm{\tiny200}}$ (8); the velocity dispersion within a $\sigma$ derived r$_{\textrm{\tiny200}}$ (9); spectroscopic completeness (to R<22) within a  $\sigma$ derived r$_{\textrm{\tiny200}}$ cut  (10), dynamical complexity within a  $\sigma$ derived r$_{\textrm{\tiny200}}$ (11);  the number of member galaxies within an X-ray derived r$_{\textrm{\tiny200}}$ (12); the radius in arcseconds of an X-ray derived r$_{\textrm{\tiny200}}$ (13); velocity dispersion within an X-ray derived r$_{\textrm{\tiny200}}$ (14); spectroscopic completeness (to R<22) within an X-ray derived r$_{\textrm{\tiny200}}$ cut (15), and the dynamical complexity within an X-ray derived r$_{\textrm{\tiny200}}$ cut (16)}
\end{deluxetable*}  

\clearpage
\pagebreak

\LongTables

\tabletypesize{\scriptsize}
\setlength{\tabcolsep}{0.05in} 
\begin{deluxetable*}{lllllcc}
\tablecaption{Basic Properties of Optically Selected Groups \label{tab:Optgrp_props1}}
\tablecolumns{7} 
\tablehead{\colhead {OID} & \colhead{RA} & \colhead{Dec} & \colhead{z} & \colhead{XID} & \colhead{X-ray Flux} & \colhead{X-ray Significance} \\  & \colhead{J2000} & \colhead{J2000} & &  & \colhead{10$^{\textrm{\tiny{-14}}}$ ergs cm$^{\textrm{\tiny{-2}}}$ s$^{\textrm{\tiny{-1}}}$} &}
\startdata
OP14h01 & 222.42588 & +9.05001 & 0.1648 & XR14h12 & 0.53$\pm$0.15 & 3.45 \\ [2pt]
OP14h08 & 222.24931 & +9.16426 & 0.2287 & \nodata &$<$ 0.18& $-$1.5 \\ [2pt]
OP14h09 & 222.22414 & +8.94529 & 0.2616 & \nodata &$<$ 0.11& 1.25 \\ [2pt]
OP14h10 & 222.38704 & +8.89722 & 0.2709 & XR14h22$^{*}$ &$<$ 0.10& $-$0.53 \\ [2pt]
OP14h11 & 222.28817 & +8.83040 & 0.2710 & XR14h25$^{*}$ &$<$ 0.11& 1.39 \\ [2pt]
OP14h15 & 222.19998 & +8.96308 & 0.3070 & \nodata &$<$ 0.13& 0.38 \\ [2pt]
OP14h16 & 222.59024 & +9.10364 & 0.3065 & \nodata &$<$ 0.31& 0.87 \\ [2pt]
OP14h19 & 222.55307 & +8.96008 & 0.3251 & \nodata &$<$ 0.18& $-$0.54 \\
\enddata    
\tablenotetext{*}{match to X-ray system but low X-ray significance using the optical center and aperture}
\tablenotetext{}{Column description: group identification number for the optical system (column 1); R.A. and Decl. of the center of the extended X-ray emission for Equinox J2000.0 (2 \& 3); spectroscopic redshift (4); X-ray group ID where there is a confident match (5); the total flux in the 0.5$-$2 keV band (6); and significance of the X-ray flux (7)}
\end{deluxetable*}

\clearpage

\tabletypesize{\tiny}
\setlength{\tabcolsep}{0.025in} 
\begin{deluxetable*}{l|ccccc|ccccc|ccccc}
  \tablecolumns{16}
  \tablecaption{Measured Properties of Optically Selected Groups \label{tab:Optgrp_props2}}
  \tablehead{\colhead{OID} & \colhead{N} & \colhead{Radius} & \colhead{$\sigma$} & completeness & dynamical & \colhead{N} & \colhead{Radius} & \colhead{$\sigma$} & completeness & dynamical & \colhead{N} & \colhead{Radius} & \colhead{$\sigma$} & completeness & dynamical \\ & \colhead{1Mpc} & \colhead{[$\arcsec$]} & \colhead{[km s$^{\textrm{\tiny{-1}}}$]} & R<22 & complexity & \colhead{r$_{\textrm{\tiny200,}\textrm{\scriptsize{$\sigma$}}}$} & \colhead{[$\arcsec$]} & \colhead{[km s$^{\textrm{\tiny{-1}}}$]} & R<22 & complexity & \colhead{r$_{\textrm{\tiny200,X}}$} & \colhead{[$\arcsec$]} & \colhead{[km s$^{\textrm{\tiny{-1}}}$]} & R<22 & complexity}
 \startdata
OP14h01 & 9 & 378.08 & 191$\pm$88 &  0.70 &-& 6 & 159.41 & 197$\pm$131 &  0.72 &-& 7 & 165.96 & 174$\pm$112 &  0.72 &-\\ [2pt]
OP14h08 & 8 & 292.19 & 159$\pm$101 &  0.59 &-& 4 & 123.00 & 204$\pm$166 &  0.65 &-&$\leq$ 4 &$<$ 115.92& 204$\pm$166 &  0.60 &-\\ [2pt]
OP14h09 & 10 & 264.17 & 168$\pm$55 &  0.58 &-& 8 & 107.71 & 201$\pm$56 &  0.40 &-&$\leq$ 8 &$<$ 100.08& 201$\pm$56 &  0.43 &-\\ [2pt]
OP14h10 & 16 & 257.58 & 367$\pm$59 &  0.72 &DS& 11 & 152.08 & 293$\pm$68 &  0.83 &-&$\leq$ 7 &$<$ 97.56 & 207$\pm$72 &  0.93 &-\\ [2pt]
OP14h11 & 18 & 257.66 & 254$\pm$42 &  0.54 &-& 15 & 134.54 & 259$\pm$50 &  0.75 &-&$\leq$ 11&$<$ 98.64 & 274$\pm$69 &  0.69 &-\\ [2pt]
OP14h15 & 7 & 235.96 & 110$\pm$88 &  0.48 &-& 6 & 39.56 & 84$\pm$110 &  0.70 &-&$\leq$ 7 &$<$ 96.84 & 110$\pm$88 &  0.65 &-\\ [2pt]
OP14h16 & 6 & 236.32 & 219$\pm$73 &  0.53 &-& 5 & 89.54 & 192$\pm$158 &  0.73 &-&$\leq$ 5 &$<$ 114.84& 192$\pm$158 &  0.75 &-\\ [2pt]
OP14h19 & 7 & 227.06 & 174$\pm$69 &  0.61 &-& 5 & 69.16 & 156$\pm$182 &  0.47 &-&$\leq$ 5 &$<$ 101.16& 156$\pm$182 &  0.48 &-\\
\enddata
\tablenotetext{}{Column description: group identification number for the optical system (column 1); number of member galaxies within 1\,Mpc (2); the radius in arcseconds of 1\,Mpc (3); the velocity dispersion within a 1\,Mpc cut (4); spectroscopic completeness (to R<22) within a 1\,Mpc cut  (5), dynamical complexity within a 1\,Mpc cut  (6); the number of member galaxies within a $\sigma$ derived r$_{\textrm{\tiny200}}$ (7); the radius in arcseconds of a $\sigma$ derived r$_{\textrm{\tiny200}}$ (8); the velocity dispersion within a $\sigma$ derived r$_{\textrm{\tiny200}}$ (9); spectroscopic completeness (to R<22) within a  $\sigma$ derived r$_{\textrm{\tiny200}}$ cut  (10), dynamical complexity within a  $\sigma$ derived r$_{\textrm{\tiny200}}$ (11);  the number of member galaxies within an X-ray derived r$_{\textrm{\tiny200}}$ (12); the radius in arcseconds of an X-ray derived r$_{\textrm{\tiny200}}$ (13); velocity dispersion within an X-ray derived r$_{\textrm{\tiny200}}$ (14); spectroscopic completeness (to R<22) within an X-ray derived r$_{\textrm{\tiny200}}$ cut (15), and the dynamical complexity within an X-ray derived r$_{\textrm{\tiny200}}$ cut (16)}
\end{deluxetable*}  

\clearpage
\pagebreak

\LongTables
\tabletypesize{\tiny}
\setlength{\tabcolsep}{0.05in} 
\begin{deluxetable*}{l|ll|lll|lll}
\tablecolumns{9}
  \tablecaption{Masses of X-ray Selected Groups \label{tab:Xgrp_masses} }
  \tablehead{\colhead{XID} & \colhead{L$_{X}$} & \colhead{M$_{X}$} & \colhead{M$_{stellar}$(1Mpc)} & \colhead{M$_{stellar}$(r$_{\textrm{\tiny200,}\textrm{\scriptsize{$\sigma$}}}$)} & \colhead{M$_{stellar}$(r$_{\textrm{\tiny200,X}}$) } & \colhead{M$_{dyn}$(1Mpc)} & \colhead{M$_{dyn}$(r$_{\textrm{\tiny200,}\textrm{\scriptsize{$\sigma$}}}$) } & \colhead{M$_{dyn}$(r$_{\textrm{\tiny200,X}}$)} \\ & \colhead{10$^{42}$ ergs s$^{-1}$} & \colhead{10$^{13}$ M$_{\sun}$} & \colhead{10$^{11}$ M$_{\sun}$} & \colhead{10$^{11}$ M$_{\sun}$} & \colhead{10$^{11}$ M$_{\sun}$} &  \colhead{10$^{13}$ M$_{\sun}$} &\colhead{10$^{13}$ M$_{\sun}$} & \colhead{10$^{13}$ M$_{\sun}$} }
\startdata
XR14h02 & 0.62$\pm$0.22 & 1.38$\pm$0.29 & 1.04$\pm^{0.55}_{0.48}$ & 0.87$\pm^{0.49}_{0.49}$ & 0.89$\pm^{0.49}_{0.48}$ & 3.170$\pm$1.001 & 0.700$\pm$0.492 & 0.869$\pm$0.612 \\ [2.15 pt] 
XR14h09 & 34.6$\pm$3.55 & 11.8$\pm$0.76 &$>$ 21.5 &$>$ 30.6 &$>$ 20.3 & 50.50$\pm$13.36 & 54.09$\pm$12.25 & 35.38$\pm$12.28 \\ [2.15 pt] 
XR14h10 & 6.55$\pm$2.64 & 3.74$\pm$0.90 &$>$ 4.69 &$>$ 4.76 &$>$ 1.31 & 14.68$\pm$11.80 & 10.32$\pm$8.297 & 8.135$\pm$10.37 \\ [2.15 pt] 
XR21h08 & 2.54$\pm$0.32 & 3.07$\pm$0.24 & 2.87$\pm^{0.78}_{0.71}$ & 2.27$\pm^{0.60}_{0.56}$ & 2.54$\pm^{0.63}_{0.54}$ & 4.815$\pm$1.415 & 2.286$\pm$0.992 & 3.078$\pm$1.017 \\ [2.15 pt] 
XR21h10 & 1.12$\pm$0.34 & 1.72$\pm$0.32 & 6.23$\pm^{3.73}_{3.46}$ & 4.14$\pm^{2.88}_{2.87}$ & 4.14$\pm^{2.88}_{2.88}$ & 4.052$\pm$1.205 & 0.602$\pm$1.395 & 0.846$\pm$1.963 \\ [2.15 pt] 
XR21h11 & 2.98$\pm$0.66 & 2.85$\pm$0.39 & 11.2$\pm^{2.97}_{2.52}$ & 10.1$\pm^{2.47}_{2.01}$ & 9.51$\pm^{2.00}_{1.83}$ & 9.766$\pm$3.945 & 8.949$\pm$3.737 & 6.474$\pm$2.641 \\ [2.15 pt] 
XR21h12 & 0.59$\pm$0.18 & 1.24$\pm$0.23 & 5.31$\pm^{0.89}_{0.84}$ & 1.06$\pm^{0.21}_{0.20}$ & 1.08$\pm^{0.19}_{0.20}$ & 4.570$\pm$1.144 & 2.705$\pm$1.601 & 2.071$\pm$1.226 \\ [2.15 pt] 
XR21h14 & 26.1$\pm$0.99 & 11.8$\pm$0.28 & 12.2$\pm^{1.93}_{1.80}$ & 21.2$\pm^{3.61}_{3.16}$ & 8.20$\pm^{1.97}_{1.74}$ & 39.24$\pm$7.647 & 40.77$\pm$7.241 & 28.41$\pm$6.542 \\
\enddata
\tablenotetext{}{\textrm{\tiny{Column description: group identification number for the X-ray system (column 1); rest-frame luminosity in the 0.1$-$2.4 keV band (2); estimates of a total mass, using X-ray luminosity as a mass proxy and a calibration of \cite{Leauthaud2010} (3); group stellar mass calculated using 1\,Mpc, r$_{\textrm{\tiny200,}\textrm{\scriptsize{$\sigma$}}}$, and r$_{\textrm{\tiny{200,X}}}$ radial cuts (4, 5, \& 6); and the dynamical (virial) mass for 1\,Mpc, r$_{\textrm{\tiny200,}\textrm{\scriptsize{$\sigma$}}}$, and r$_{\textrm{\tiny{200,X}}}$ radial cuts (7, 8, \& 9).}}}
\end{deluxetable*}  

%\clearpage
%\pagebreak
%\LongTables

\tabletypesize{\scriptsize}
\setlength{\tabcolsep}{0.05in} 
\begin{deluxetable}{l|ll|lll|lll}
\tablecolumns{9}
  \tablecaption{Masses of Optically Selected Groups \label{tab:Optgrp_masses}}
  \tablehead{\colhead{OID} & \colhead{L$_{X}$} & \colhead{M$_{X}$} & \colhead{M$_{stellar}$(1Mpc)} & \colhead{M$_{stellar}$(r$_{\textrm{\tiny200,}\textrm{\scriptsize{$\sigma$}}}$)} & \colhead{M$_{stellar}$(r$_{\textrm{\tiny200,X}}$) } & \colhead{M$_{dyn}$(1Mpc)} & \colhead{M$_{dyn}$(r$_{\textrm{\tiny200,}\textrm{\scriptsize{$\sigma$}}}$) } & \colhead{M$_{dyn}$(r$_{\textrm{\tiny200,X}}$)} \\ & \colhead{10$^{42}$ ergs s$^{-1}$} & \colhead{10$^{13}$ M$_{\sun}$} & \colhead{10$^{11}$ M$_{\sun}$} & \colhead{10$^{11}$ M$_{\sun}$} & \colhead{10$^{11}$ M$_{\sun}$} &  \colhead{10$^{13}$ M$_{\sun}$} &\colhead{10$^{13}$ M$_{\sun}$} & \colhead{10$^{13}$ M$_{\sun}$} }
\startdata
OP14h01 & 0.62$\pm$0.18 & 1.34$\pm$0.23 & 0.87$\pm^{0.41}_{0.33}$ & 0.52$\pm^{0.28}_{0.28}$ & 0.53$\pm^{0.30}_{0.27}$ & 2.565$\pm$2.375 & 1.153$\pm$1.530 & 0.928$\pm$1.200  \\ [2.25 pt]  
OP14h08 &$<$ 0.46&$<$ 1.05& 0.77$\pm^{0.21}_{0.19}$ & 0.46$\pm^{0.13}_{0.12}$ &$\leq$ 0.57$\pm^{0.10}_{0.10}$ & 1.781$\pm$2.253 & 1.230$\pm$2.007 & 1.159$\pm$1.891  \\ [2.25 pt]  
OP14h09 &$<$ 0.40&$<$ 0.95& 1.32$\pm^{0.49}_{0.46}$ & 1.37$\pm^{0.45}_{0.38}$ &$\leq$ 1.35$\pm^{0.42}_{0.39}$ & 1.983$\pm$1.304 & 1.154$\pm$0.641 & 1.073$\pm$0.596  \\ [2.25 pt]  
OP14h10 &$<$ 0.42&$<$ 0.96& 3.74$\pm^{1.19}_{1.08}$ & 2.90$\pm^{1.16}_{1.03}$ &$\leq$ 2.33$\pm^{1.06}_{1.15}$ & 9.425$\pm$2.824 & 3.540$\pm$1.652 & 1.142$\pm$0.800  \\ [2.25 pt]  
OP14h11 &$<$ 0.44&$<$ 0.99& 2.80$\pm^{0.50}_{0.41}$ & 2.24$\pm^{0.41}_{0.35}$ &$\leq$ 1.71$\pm^{0.42}_{1.71}$ & 4.509$\pm$1.503 & 2.453$\pm$0.952 & 2.011$\pm$1.012  \\ [2.25 pt]  
OP14h15 &$<$ 0.67&$<$ 1.27& 3.64$\pm^{1.71}_{1.49}$ & 1.40$\pm^{0.34}_{0.35}$ &$\leq$ 1.26$\pm^{0.45}_{0.36}$ & 0.859$\pm$1.377 & 0.084$\pm$0.219 & 0.352$\pm$0.565  \\ [2.25 pt]  
OP14h16 &$<$ 1.50&$<$ 2.12& 1.67$\pm^{0.75}_{0.59}$ & 1.44$\pm^{0.64}_{0.73}$ &$\leq$ 1.41$\pm^{0.77}_{0.72}$ & 3.367$\pm$2.261 & 0.975$\pm$1.605 & 1.250$\pm$2.059  \\ [2.25 pt]  
OP14h19 &$<$ 1.04&$<$ 1.65& 4.20$\pm^{1.13}_{1.11}$ & 3.15$\pm^{0.51}_{0.51}$ &$\leq$ 2.67$\pm^{0.62}_{0.64}$ & 2.124$\pm$1.689 & 0.517$\pm$1.213 & 0.756$\pm$1.774  \\
\enddata
\tablenotetext{}{\textrm{\tiny{Column description: group identification number for the optical system (column 1); rest-frame luminosity in the 0.1$-$2.4 keV band (2); estimates of a total mass, using X-ray luminosity as a mass proxy and a calibration of \cite{Leauthaud2010} (3); group stellar mass calculated using 1\,Mpc, r$_{\textrm{\tiny200,}\textrm{\scriptsize{$\sigma$}}}$, and r$_{\textrm{\tiny{200,X}}}$ radial cuts (4, 5, \& 6); and the dynamical (virial) mass for 1\,Mpc, r$_{\textrm{\tiny200,}\textrm{\scriptsize{$\sigma$}}}$, and r$_{\textrm{\tiny{200,X}}}$ radial cuts (7, 8, \& 9).}}}
\end{deluxetable}  

\clearpage
\pagebreak

%% End of file 
\end{document}